%% file: Reliability_New_ArXiv.tex
\documentclass[onecolumn,12pt]{IEEEtran}

\usepackage{amssymb,amsfonts,amsmath,amstext,mathrsfs}
\usepackage{latexsym}
\usepackage{eufrak}
\usepackage{epsfig,psfrag,color,graphics,epsf}
\usepackage{enumerate,cite}
\usepackage{bbm}
\usepackage{stmaryrd}
\usepackage{soul}
\usepackage{tcolorbox}

\usepackage{amsmath}
\usepackage{amssymb}
\usepackage{amsfonts}
\usepackage{amsthm}
\usepackage{graphicx}
\usepackage{float}
\usepackage{graphics}
\usepackage{graphicx}
\usepackage{euscript}
\usepackage{psfrag}

\usepackage{standalone}
\usepackage{tikz,pgfplots}

\DeclareMathOperator*{\argmax}{arg\,max}

\DeclareMathOperator*{\argsup}{arg\,sup}

\newtheorem{theorem}{Theorem}
\newtheorem{lemma}{Lemma}

\newtheorem{proposition}{Proposition}

\newtheorem{corollary}{Corollary}
\newtheorem{definition}{Definition}

\newcommand{\dotleq}{%
\DOTSB\mathrel{\mathop{\kern0pt \leq}\limits^{\textstyle.}}}
\newcommand{\dotgeq}{%
\DOTSB\mathrel{\mathop{\kern0pt \geq}\limits^{\textstyle.}}}

\newcommand{\reals} {\mathbb{R}}

\newcommand{\beq} {\begin{equation}}
\newcommand{\eeq} {\end{equation}}
\newcommand{\beqa} {\begin{align}}
\newcommand{\eeqa} {\end{align}}

\newcommand{\indicator}{\mathbbm{1}}

\newcommand {\bv} {\boldsymbol{v}}

\newcommand {\bx} {\boldsymbol{x}}
\newcommand {\by} {\boldsymbol{y}}
\newcommand {\bz} {\boldsymbol{z}}

\newcommand {\bX} {\boldsymbol{X}}
\newcommand {\bY} {\boldsymbol{Y}}
\newcommand {\bZ} {\boldsymbol{Z}}

\newcommand{\calA}{{\mathcal A}}
\newcommand{\calB}{{\mathcal B}}
\newcommand{\calC}{{\mathcal C}}

\newcommand{\calE}{{\mathcal E}}

\newcommand{\calG}{{\mathcal G}}

\newcommand{\calK}{{\mathcal K}}
\newcommand{\calL}{{\mathcal L}}

\newcommand{\calP}{{\mathcal P}}
\newcommand{\calQ}{{\mathcal Q}}

\newcommand{\calS}{{\mathcal S}}
\newcommand{\calT}{{\mathcal T}}
\newcommand{\calU}{{\mathcal U}}
\newcommand{\calV}{{\mathcal V}}
\newcommand{\calW}{{\mathcal W}}
\newcommand{\calX}{{\mathcal X}}
\newcommand{\calY}{{\mathcal Y}}
\newcommand{\calZ}{{\mathcal Z}}

\newcommand{\EE}{{\mathbb E}}

\allowdisplaybreaks
\IEEEoverridecommandlockouts

\begin{document}



\title{An Upper Bound on the Reliability Function of the DMC
}

\author{\IEEEauthorblockN{Anelia Somekh-Baruch\\}
\IEEEauthorblockA{\tt{Faculty of Engineering\\ Bar-Ilan University \\
Ramat-Gan, Israel}\\
\tt{Email: somekha@biu.ac.il}
}
\thanks{
This work was supported by the Israel Science Foundation (ISF) grant no.\ 631/17.
}}

\maketitle
\begin{abstract}

We derive a new upper bound on the reliability function for channel coding over discrete memoryless channels. Our bounding technique relies on two main elements: (i) adding an auxiliary genie-receiver that reveals to the original receiver a list of codewords including the transmitted one, which satisfy a certain type property, and (ii) partitioning (most of) the list into subsets of codewords that satisfy a certain pairwise-symmetry property, which facilitates lower bounding of the average error probability by the pairwise error probability within a subset. We compare the obtained bound to the Shannon-Gallager-Berlekamp straight-line bound, the sphere-packing bound, and  an amended version of Blahut's bound. Our bound is shown to be at least as tight for all rates, with cases of stricter tightness in a certain range of low rates, compared to all three aforementioned bounds. Our derivation is performed in a unified manner which is valid for any rate, as well as for a wide class of additive decoding metrics, whenever the corresponding zero-error capacity is zero. We further present a relatively simple function that may be regarded as an approximation to the reliability function in some cases. We also present a dual form of the bound, and discuss a looser bound of a simpler form, which is analyzed for the case of the binary symmetric channel with maximum likelihood decoding. 
\end{abstract}

\section{Introduction}

It is well known that the average error probability in coding for the discrete memoryless channel (DMC) can be made to vanish exponentially fast with the code block-length $n$, if the rate of the code is strictly lower than the channel capacity.
The reliability function signifies exactly how fast the exponential decay can be, as a function of the code rate $R$.  
The problem of finding a {\it single-letter} expression for the reliability function has been studied extensively since the early days of information theory (see, e.g., \cite{Gallager68,CsiszarKorner81,SHANNONGallagerBerlekamp1967522,shannonGallagerBerlekamp1967lower,BlahutComposition}). 

For a certain range of high rates, the random-coding lower bound coincides with the sphere-packing upper bound \cite{Gallager68,CsiszarKorner81}, establishing the exact value of the reliability function curve above a certain critical rate. 
The exact value of the reliability function for zero-rate codes (at $R=0^+$) is also known, and was characterized by Shannon, Gallager and Berlekamp \cite{SHANNONGallagerBerlekamp1967522} who obtained a tight upper bound that matches the expurgated lower bound at zero rate \cite{Gallager68}, for channels whose zero-error capacity equals zero. 
Another important result of \cite{SHANNONGallagerBerlekamp1967522}, which holds for the same class of channels, 
is referred to as the straight-line upper bound on the reliability function at low rates, which is obtained by the tangential line to the sphere packing bound curve that meets the $R=0$ axis at the zero-rate reliability point.

In \cite{BlahutComposition} an upper bound on the reliability
function 
was proposed, 
that contained a gap in the proof, which was revisited 
in \cite{BondaschiDalai2022}. The correction was based on
Blinovsky’s \cite{Blinovsky2002} idea of using a
Ramsey-theoretic result by Koml\'{o}s \cite{komlos1990strange}.
This yielded an 
 upper bound applicable to 
general DMCs. 

For certain special cases of channels, tighter upper bounds on the reliability function have been derived sometimes enlarging the range of rates for which it is known, see, e.g., \cite{burnashev2016reliability}. 
However, 
deriving a single-letter formula for the reliability function for a certain range of low rates remains an open problem, and so is the question of understanding the structure of low rate good codes. 
For a survey of the subject of the channel coding reliability function see \cite{Haroutunian_Reliability_FnT2008} and references therein.

Certain works have extended the study of the reliability function of the DMC to the more general setup, where the decoding metric is not necessarily optimal, a.k.a.\ mismatched decoding (see
\cite{CsiszarKorner81graph,SomekhBaruch_mismatchachievableIT2014,ScarlettPengMerhavMartinezGuilleniFabregas_mismatch_2014_IT,ScarlettGuilleniFabregasSomekhBaruchMartinez2020,BondaschiGuilleniFabregasDalai-IT2021,KangarshahiGuilleniFabregas2021spherepackingArXiv,SomekhBaruchArxiv_16March2022}). 

In this paper, we use a framework for proving converse theorems that we developed
in a series of recent works (\hspace{1sp}\cite{SomekhBaruch2020singleletter_part1,SomekhBaruch_ITW_2020,SomekhBaruch_ISIT2021,SomekhBaruchArxiv_16March2022}).
This method relies on extending the single-user channel to a two-output (broadcast) channel, and simultaneously transmitting the signal to both receivers. 
This technique, which is referred to as multicast transmission with an auxiliary receiver, was used to derive several bounds on the mismatch capacity of and the mismatched reliability function of the DMC. 
A refinement of this method was introduced in \cite{SomekhBaruchArxiv_16March2022} which yielded the tightest converse bounds to date for these two problems. 
The main idea of the proof is based on the fact that the auxiliary receiver serves as a genie that reveals to the original receiver a suitably-chosen narrowed list of codewords, that includes the transmitted one, satisfying certain type-based properties. 

We combine this proof technique with an additional element. We partition (most of) the list into disjoint sets (sub-lists) of codewords that satisfy a certain pairwise-symmetry property. This enables to obtain a bound on the reliability function, with or without mismatch, in terms of the average pairwise error probability within a sub-list, that can be bounded using 
the aforementioned Blinkovsky's idea \cite{Blinovsky2002} of using the 
Ramsey-theoretic result by Koml\'{o}s, and 
Plotkin's counting trick similar to \cite{BondaschiDalai2022}. 
The resulting bound is tight at $R=0^+$; that is, recovers the result of \cite{SHANNONGallagerBerlekamp1967522} for a matched metric, and that of \cite{BondaschiGuilleniFabregasDalai-IT2021} for a mismatched metric. It is also at least as tight as the sphere-packing bound and the bound of \cite{SomekhBaruchArxiv_16March2022} in the mismatched case, and is strictly tighter than the straight-line bound of
\cite{SHANNONGallagerBerlekamp1967522}, at least in certain cases. 
We also show how the 
bound of \cite{BondaschiDalai2022} which contained a multi-letter ingredient, can be transformed 
to yield a purely single-letter upper bound of a similar form, and demonstrate that in the general case, our bound is strictly tighter also compared to the latter.

This paper is organized as follows: 
In Section \ref{sc: A Formal Statement of the Problem} we present a statement of the problem. 
In Section \ref{cs: Previous Bounds}, we discuss previous bounds and related results, which are most relevant to our work. 
Section \ref{sc: NewResults} summarizes our main results. 
Section \ref{sc: Proof of Exponent Theorem } is dedicated to the proof of the main theorem regarding the basic bound on the reliability function.
Section \ref{sc: Conclusion} discusses some concluding remarks.
Proofs of additional results and lemmas appear in the appendix.

\section{A Formal Statement of the Problem}\label{sc: A Formal Statement of the Problem}

Consider transmission over a memoryless channel described by a conditional probability rule $W(y|x)$, with input $x\in\calX$ and output $y\in\calY$ finite alphabets $\calX$ and $\calY$; in particular, $W(y|x)$ is a conditional probability mass function. We define 
$
W^n(\by|\bx) = \prod_{k=1}^n W(y_k|x_k)
$
for input/output sequences $\bx = (x_1,\dotsc,x_n)\in\calX^n$ and $\by= (y_1,\dotsc,y_n)\in\calY^n$. The corresponding random variables are denoted by $\bX$ and $\bY$.

An encoder maps a message $m\in \{1,\dotsc,\mathbb{M}_n\}$ to a channel input sequence $\bx_m\in\calX$, 
where the number of messages is denoted by $\mathbb{M}_n$.
The message, represented by the random variable $M$, is assumed to take values in $\{1,\dotsc,\mathbb{M}_n\}$ equi-probably. This mapping induces an $(n,\mathbb{M}_n)$-codebook $\calC_n=\{\bx_1,\dotsc,\bx_{\mathbb{M}_n}\}$ with rate $R_n=\frac{1}{n}\log \mathbb{M}_n$.

Upon observing the channel output $\by$, the decoder, who wishes to minimize the average error probability, can use the optimal ML decoding rule
\beq
\widehat m = \argmax_{m\in\{1,\dotsc,\mathbb{M}_n\}} \,W^n(\by|\bx_m)= \argmax_{m\in\{1,\dotsc,\mathbb{M}_n\}}\sum_{ i=1}^n q_{\mbox{\tiny{ML}}} (x_{m,i},y_i),\label{eq:ML decoder}
\eeq
where $x_{m,i}$ denotes the $i$-symbol of the $m$-th codeword, $\bx_m$, and 
\begin{flalign}
q_{\mbox{\tiny{ML}}}(x,y) \triangleq \log W(y|x)
\end{flalign} is referred to as the ML metric. 

This can be generalized to a sub-optimal (mismatched) decoding rule \cite[Ch.~2]{CsiszarKorner81} :
\beq
\widehat m = \argmax_{m\in\{1,\dotsc,\mathbb{M}_n\}} \,\sum_{i=1}^n q(x_{i,m},y_i), \label{eq:decoder}
\eeq
defined by a {\em single-letter} mapping $q:\calX\times\calY\rightarrow \reals\cup\{-\infty\}$. 
Throughout the paper we assume that ties are broken uniformly between the maximizers. Further we assume that
\begin{flalign}
W(y|x)>0\Rightarrow q(x,y)>-\infty\label{eq: positive capacity}
\end{flalign}
as otherwise without loss of generality any symbol $x$ for which there exists $y$ satisfying $W(y|x)>0$ and $q(x,y)=-\infty$ should not be used in the codebook, since any codeword containing this symbol results in erroneous decoding with probability $W(y|x)$, which is bounded away from zero, regardless of the block length.

Denote the random variable corresponding to the decoded message by $\widehat M_q(\bY)$, and the average error probability as
$
P_e(W,\calC_n,q) = \Pr \bigl[\widehat M_q(\bY) \neq M\bigr]
$.

A rate $R$ is said to be achievable with decoding metric $q$ if there exists a sequence of codebooks $\calC_n$, $n=1,2,...$ such that $\frac{1}{n}\log|\calC_n|\geq R$ and $\lim_{n\rightarrow \infty}P_e(W,\calC_n,q)=0$. 
The channel capacity w.r.t.\ metric $q$, denoted $C_q(W)$, is defined as the supremum of achievable rates, and is referred to as the mismatch capacity. 

Evidently, Shannon's channel capacity $C(W)$ can be viewed in fact as the channel capacity w.r.t.\ the metric $q_{\mbox{\tiny{ML}}}$; that is, 
\begin{flalign}
C(W)=C_q(W)|_{q=q_{\mbox{\tiny{ML}}}=\log W}.
\end{flalign}

Other quantities that concern channel coding with zero-error and are relevant to our analysis are defined next. 
The zero-error capacity of channel $W$, denoted $C_0(W)$, is defined as the supremum of rates for which there exists a sequence of codebooks $\calC_n$, $n=1,2,...$, such that the average error probability with maximum ML decoding is equal to zero. 
The zero-error capacity with decoding metric $q$, $C_{0,q}(W)$, is defined as the supremum of rates for which there exists a sequence of codebooks $\calC_n$, $n=1,2,...$, such that $P_e(W,\calC_n,q)=0$ where ties must be broken uniformly among the minimizers.

In \cite{BondaschiGuilleniFabregasDalai-IT2021}, conditions for $C_{0,q}(W)=0$ were derived for a multiplicative metric, which can be phrased as follows for the case of an additive metric $q$:
\begin{flalign}\label{eq: C 0 conditions}
C_{0,q}(W_{Y|X})=0\mbox{  iff }\forall (x,\widetilde{x}),\; \max_{y:\; W(y|\widetilde{x})>0}[q(x,y)-q(\widetilde{x},y)]+ \max_{y:\; W(y|x)>0}[q(\widetilde{x},y)-q(x,y)]\geq 0 .
\end{flalign}

A rate-exponent pair $(R,E)$ is said to be achievable 
    for channel $W$ with decoding metric $q$ if there exists a sequence of codebooks $\calC_n$, $n=1,2,...$ such that 
    for all $n$, $\frac{1}{n}\log|\calC_n|\geq R$ and 
    \begin{flalign}
        \liminf_{n\to\infty} \, -\,\frac{1}{n}\log P_e(W,\calC_n,q)\geq E. 
    \end{flalign}
    Equivalently, we say that $E$ is an achievable error exponent at rate $R$ if $(R,E)$ is an achievable rate-exponent pair.

The reliability function of the channel with decoding metric $q$ is the supremum of achievable error exponents as a function of the code rate, and is denoted by $E^q(R,W)$. The reliability function with the optimal ML decoding metric $q_{\mbox{\tiny{ML}}}(x,y)=\log W(y|x)$ is denoted $E(R,W)$; i.e., 
\begin{flalign}
E(R,W)&\triangleq \left.E^q(R,W)\right|_{q=q_{\mbox{\tiny{ML}}}}.\label{eq: afuddiadvadsvudgfuvf}
\end{flalign}

We next introduce some notation.  
For a given sequence $\bx \in \calX^n$, where $\calX$ is a finite alphabet,  $\widehat{P}_{\bx}$ denotes the empirical distribution on $\calX$ extracted from $\bx$; in other words, the vector $\{ \widehat{P}_{\bx} (x), x\in\calX\}$, where $ \widehat{P}_{\bx} (x)$ is the relative frequency of the symbol $x$ in the vector $\bx$. The type-class of $\bx$ is the set of $\bx'\in\calX^n$ such that $\widehat{P}_{\bx'}=\widehat{P}_{\bx}$, which is denoted $\calT_n(\widehat{P}_{\bx})$. 

Define the highest achievable exponent with $P$ constant composition codebooks of block length $n$ and $q$-decoding as
\begin{flalign}\label{eq: afuivudgfuvf}
e_n^q(R,P,W)&\triangleq 
\max_{\calC_n\subseteq\calT_n(P):\; |\calC_n|\geq e^{nR}}
-\frac{1}{n}\log P_e(q,W,\calC_n).
\end{flalign}
Using standard arguments that follow from the fact that the number of type-classes grows polynomially with $n$, it can be shown that
\begin{flalign}\label{eq: afuddivudgfuvf}
E^q(R,W)=\liminf_{n\rightarrow \infty}\max_{P_n\in\calP_n(\calX)} e_n^q(R,P_n,W).
\end{flalign}

\section{Previous Bounds and Related Results}\label{cs: Previous Bounds}

In this section we mention previous bounds and results that are most relevant to this work. 
For a comprehensive survey 
the reader is referred to the additional aforementioned works and references therein.

The classical sphere-packing {\it upper} bound on $E(R,W)$ is given by:
\begin{flalign}\label{eq: E_esp}
E_{sp}(R,W)=& \max_{P\in\calP(\calX)} E_{sp}(R,P,W)
\end{flalign}
where $\calP(\calX)$ denotes the set of all probability distributions on $\calX$ and
\begin{flalign}
  E_{sp}(R,P,W)=& \min_{P_{Y|X}:\; I(P\times P_{Y|X})\leq R }D(P_{Y|X}\|W|P), 
\end{flalign}
which holds for the ML metric, and thus also for any other metric.

The expurgated and random coding\footnote{Here the notion ``random coding" refers to a scheme in which the codewords are drawn independently and uniformly within a single type-class. } {\it lower} bounds on the reliability function for coding with decoding metric $q$, $E^q(R,W)$ (\hspace{1sp}\cite{Gallager68,CsiszarKorner81graph,RGV-IT2019}), are given by: 
\begin{flalign}
&E^q_{ex}(R,W)=\max_{P\in\calP(\calX)}\min_{ \substack{V:
V_X=V_{\widetilde{X}}=P,\\ \EE q(\widetilde{X},Y)\geq \EE q(X,Y),\\ I(\widetilde{X};X)\leq R} }D(V_{Y|X}\|W|P)+ I(\widetilde{X};Y,X)-R\\
&E^q_{r}(R,W)=\max_{P\in\calP(\calX)}\min_{ \substack{V:
V_X=V_{\widetilde{X}}=P,\\ \EE q(\widetilde{X},Y)\geq \EE q(X,Y)} }D(V_{Y|X}\|W|P)+\left|I(\widetilde{X};Y,X)-R\right|_+,
\end{flalign}
respectively, where $|t|_+\triangleq\max\{0,t\}$.

Csisz\'{a}r and K\"{o}rner's \cite{CsiszarKorner81} derived the following {\it lower} bound 
\begin{flalign}
E^q_{\mbox{\tiny{CK}}}(R,W)  &= \max_{P\in\calP(\calX)}E^q_{\mbox{\tiny{CK}}}(R,P,W)\label{eq: E_ex dfn C and K 22 22 udgfugeug}\\
E^q_{\mbox{\tiny{CK}}}(R,P,W)&=\min_{\substack{V: V_X=V_{\widetilde{X}}=P,\\ \EE q(\widetilde{X},Y)\geq \EE q(X,Y),\\ I(\widetilde{X};X)\leq R}}D(V_{Y|X}\|W|P)+\left|I(\widetilde{X};Y,X)-R\right|_+,\label{eq: E_ex dfn C and K 22 22 udgfugeug2}
\end{flalign}
 which 
is at least as tight as the maximum between $E^q_{ex}(R,W)$ and the random coding bound.

For the zero-rate case it becomes \begin{flalign}
&E^q_{\mbox{\tiny{CK}}}(0^+,W)=E^{q}_{ex}(0^+,W)\triangleq \max_{P\in\calP(\calX)}\;
\min_{ \substack{ P_{\widetilde{X}Y|X}:\;I(\widetilde{X};X)=0,\\
P_{\widetilde{X}}=P,\\
\EE q(\widetilde{X},Y)\geq \EE q(X,Y)
} } 
D(P_{Y|X}\|W|P_X)+I(\widetilde{X};Y|X)
.
 \end{flalign}
Shannon, Gallager and Berlekamp \cite{SHANNONGallagerBerlekamp1967522} showed that for $q=q_{\mbox{\tiny{ML}}}$, the bound is tight, that is;
\begin{flalign}
E(0^+,W)&= E_{ex}(R,W)\triangleq \left. E^q_{ex}(R,W)\right|_{q=q_{\mbox{\tiny{ML}}}},
\end{flalign}
for channels satisfying $C_0(W)=0$. 

Another important result of \cite{SHANNONGallagerBerlekamp1967522} is the derivation of the following 
combined upper bound 
\begin{flalign}
&E_{sl-sp}(R,W)=\nonumber\\
&\left\{\begin{array}{ll}
-\frac{[E(0^+,W)-E_{sp}(R_{W}^*,W)]}{R_{W}^*} R+E(0^+,W)& R<R_{W}^*\\
E_{sp}(R,W) & R>R_{W}^*
\end{array}\right.\label{eq: SLLSP}
\end{flalign} 
where $R_{W}^*$ signifies the $R$-coordinate of the point at which the tangential line from the point $(0,E(0^+,W))$ meets the $E_{sp}(\cdot ,W)$ curve in the $(R,E)$-plane. The linear part of the curve; i.e., for $R\in(0, R_{W}^*)$, is referred to as the straight-line bound.

As mentioned above, in \cite{BondaschiDalai2022}, a correction to the proof of \cite{BlahutComposition} was suggested,
which yielded an amended upper bound (see \cite[Eq.\ (89)]{BondaschiDalai2022} applied to the case $L=1$), 
that can be expressed as:
\begin{flalign}
&\max_{P\in\calP(\calX)}\min_{\substack{\hat{P}_{XV}\in\calP_n(\calX\times\calV):\\
\hat{P}_X=P,\\ I(X;V)\leq R}}
\sum_v \hat{P}_V(v)\mathfrak{C}\left(\sum_{(x,\widetilde{x})\in\calX^2} 
\hat{P}_{X|V}(x|v)\hat{P}_{X|V}(\widetilde{x}|v)\mu_{x,\widetilde{x}}\right) \\
&\mu_{x,\widetilde{x}}\triangleq 
\log\frac{1}{\sum_y\sqrt{W(y|x)W(y|\widetilde{x})}}
,\label{eq: 16sfhivdhv}
\end{flalign}
where $\calP_n(\calA)$ denotes the set of empirical distributions of order $n$ on alphabet $\calA$, $V$ is a random variable of finite alphabet $\calV$, and for a probability vector $(\alpha_1,...,a_k)$ satisfying $\forall i, \alpha_i\geq 0$ and $\sum_{i=1}^k\alpha_i=1$, the notation $\mathfrak{C}\left(\sum_{i\in\{1,...,k\}} 
\alpha_i f(i)\right)$ stands for the upper concave envelope of $\sum_{i\in\{1,...,k\}} 
\alpha_i f(i)$ as a function of $(\alpha_1,...,a_k)$. 
Note that 
for a fixed rational $P\in\calP(\calX)$, as is, the expression inside the maximization involves a minimization over $\calP_n(\calX\times\calV)$ and as such, does not constitute a single-letter expression. Further, strictly speaking, 
any irrational choice of $P\in\calP(\calX)$ in the maximization of (\ref{eq: 16sfhivdhv}) results in the set $\{\hat{P}_{XZ}\in\calP_n(\calX\times\calZ):\;
\hat{P}_X=P\}$ being empty.

In the mismatched case, we recently obtained results \cite{SomekhBaruchArxiv_16March2022} which provide the mismatch counterpart of the sphere-packing bound:
\begin{flalign}
 E_{sp}^q(R,P,W)
&\triangleq 
\min_{\substack{P_{YZ|X}\in \calW_q(P),\;  I(X;Z)\leq R\\
}}
D(P_{Y|X}\|W|P),\label{eq: sp 1}\end{flalign}
where
\begin{flalign}
\calW_q(P_X)=\bigg\{
P_{YZ|X}:\; \min_{\substack{V_{U\widetilde{X}XZ}= V_{UX\widetilde{X}Z}:\\ V_{XZ}=P_{XZ}\\
\widetilde{X}-(U,Z)-X
 }}\EE_{V_{U\widetilde{X}XZ}P_{Y|XZ}} q(\widetilde{X},Y)\geq \EE q(X,Y) \bigg\},
\label{eq: calW q dfnasdfionidaf;i;}
\end{flalign}
and where $U$ is an auxiliary random variable
with alphabet size $|\calU|\leq |\calX|^2|\calZ|$, and the condition $V_{U\widetilde{X}XZ}= V_{UX\widetilde{X}Z}$ signifies that for all $(u,x_1,x_2,z)$, $V_{U\widetilde{X}XZ}(u,x_1,x_2,z)= V_{U\widetilde{X}XZ}(u,x_2,x_1,z)$.

In \cite{BondaschiGuilleniFabregasDalai-IT2021}, the expurgated bound $E^q_{ex}(R,W)$ was shown to be tight at $R=0^+$ also in the mismatched case for the wide class of balanced channel-metric pairs defined as follows:
\begin{definition}(\hspace{1sp}\cite{BondaschiGuilleniFabregasDalai-IT2021}) \label{df: balanced} 
A discrete memoryless channel $W_{Y|X}$ and a
decoding metric $q$ form a {\it balanced} pair if $C_{0,q}(W)=0$ and
\begin{flalign}\label{eq: balanced C 0 conditions}
&\max_{y:\; W(y|\widetilde{x})>0}[q(x,y)-q(\widetilde{x},y)]= \min_{y:\; W(y|x)>0}[q(x,y)-q(\widetilde{x},y)]\Rightarrow \nonumber\\
&\max_{\substack{y:\; W(y|x)+ W(y|\widetilde{x})>0,\\ q(x,y) \neq -\infty ,\; q(\widetilde{x},y) \neq -\infty }}
[q(x,y)-q(\widetilde{x},y)]=
\min_{\substack{y:\; W(y|x)+ W(y|\widetilde{x})>0,\\ q(x,y) \neq -\infty ,\; q(\widetilde{x},y) \neq -\infty }}
[q(x,y)-q(\widetilde{x},y)]
\end{flalign}
\end{definition}
As noted in \cite{BondaschiGuilleniFabregasDalai-IT2021}, 
all channels and decoding metrics such that $C_{0,q}(W)=0$ and 
\begin{flalign}
W(y|x) > 0\; \Leftrightarrow  q(x,y) >-\infty
\end{flalign}
are balanced, and obviously form a very important
wide class of channel-metric pairs, which includes the ML metric case of channels for which $C_0(W)=0$.

\section{Main Results}\label{sc: NewResults}

This section is devoted to presenting the new upper bound on $E^q(R,W)$. 
The main theorem and the outline of its proof are presented in Section \ref{sc: The Main Theorem}. 
We further present a dual form of the bound in Section \ref{sc: Dual Form of the Bound}, and show in Section \ref{sc: Comparison to Previous Results} that our bound is at least as tight as previous results, and can be strictly tighter for a certain range of low rates. 
We further introduce a looser bound in Section \ref{sc: Introduction of a Looser Bound and an Example: the BSC with ML Decoder}, and simplify it for the case of the BSC with ML decoding for which we present the resulting numerical results. We conclude in Section \ref{sc: Approximating Function - a Function that Lies Between our Upper Bound and the Lower Bound}, by presenting a function that lies between our upper bound and the lower bound $E_{\mbox{\tiny{CK}}}(R,P,W)$ that may serve as an approximation to the reliability function, when the gap between the bounds is small.

\subsection{The Main Theorem}\label{sc: The Main Theorem}
Let $W_{YZ|X}$ be a DMC from $\calX$ to $\calY\times \calZ$. Throughout this paper, we adopt the shorthand notation that $W$ without subscript signifies the original single-user channel $W_{Y|X}$. Whenever we refer to other marginal distributions; i.e., $W_{Y|XZ}$, $W_{Z|XY}$ or $W_{YZ|X}$, the subscript is mentioned explicitly.

Let $U$ denote a random variable over alphabet $\calU$ of cardinality $|\calU|\leq |\calX|^2|\calZ|+1$. 
Consider the following function
\begin{flalign}
&\overline{E}^{q}(R,P,W_{YZ|X})\triangleq \min_{\substack{P_{XZ}:\\
P_X=P,\\  I(X;Z)\leq R }}
\overline{E}^{q}(P_{XZ},W_{YZ|X}),
\label{eq: afouhvuidfhv}
\end{flalign}
where
\begin{flalign}
&\overline{E}^{q}(P_{XZ},W_{YZ|X})\triangleq \max_{\substack{P_{U\widetilde{X}|XZ}:\\ \widetilde{X}-(U,Z)-X\\
P_{\widetilde{X}ZU}=P_{XZU}
}}
\min_{ \substack{ P_{Y|UXZ\widetilde{X}}:\\ U-(\widetilde{X},X,Z)-Y,\\
\EE q(\widetilde{X},Y)\geq \EE q(X,Y)
} } 
D(P_{YZ|X}\|W_{YZ|X}|P)+I(\widetilde{X};Y|X,Z). 
\label{eq: afouhvuidfhvadvuihfdih}
\end{flalign}

For the ML decoding metric denote
\begin{flalign}
\overline{E}(R,P,W_{YZ|X})= \left. \overline{E}^{q}(R,P,W_{YZ|X})\right|_{q=q_{\mbox{\tiny{ML}}}}.
\end{flalign}

Note that the maximization and minimization in (\ref{eq: afouhvuidfhvadvuihfdih}) are in fact over $P_{U|XZ}$ and $ V_{Y|XZ\widetilde{X}}$, respectively, due to the constraints $\widetilde{X}-(U,Z)-X$, $P_{\widetilde{X}ZU}=P_{XZU}$, and $U-(\widetilde{X},X,Z)-Y$, and thus, 
$\overline{E}^{q}(P_{XZ},W_{YZ|X})$ can also be expressed as
\begin{flalign}
&\overline{E}^{q}(P_{XZ},W_{YZ|X})\nonumber\\
&=
\max_{P_{U|XZ}}
\min_{ \substack{ P_{Y|XZ\widetilde{X}}:\\ 
\EE q(\widetilde{X},Y)\geq \EE q(X,Y)
} } 
D(P_{Z|X}\|W_{Z|X}|P)+\phi(P_{XZU},P_{Y|XZ\widetilde{X}},W_{Y|XZ}),\label{eq: afouhvuidfhvadv}
\end{flalign}
where $P_{XZU}=P_{XZ}\times P_{U|XZ}$, and $\phi(P_{XZU},P_{Y|XZ\widetilde{X}},W_{Y|XZ})$ is the following divergence
\begin{flalign}
\phi(P_{XZU},P_{Y|XZ\widetilde{X}},W_{Y|XZ})&\triangleq \sum_{u,z,x,\widetilde{x},y}P_{XZU}(x,z,u)P_{X|UZ}(\widetilde{x}|u,z)
P_{Y|XZ\widetilde{X}}(y|x,z,\widetilde{x})\log \frac{P_{Y|XZ\widetilde{X}}(y|x,z,\widetilde{x})}{W_{Y|XZ}(y|x,z)}.
\end{flalign}

Due to (\ref{eq: afuddivudgfuvf}), our first result concerning the reliability function is presented in terms of an upper bound on $e_n^q(R,P,W)$. Recall Definition \ref{df: balanced} (see (\ref{eq: balanced C 0 conditions})) of a balanced channel-metric pair. 
\begin{theorem}\label{th: Main Theorem }
If $(q,W_{Y|X})$ is balanced, then for any
$|\calZ|<\infty$, $\epsilon>0$ and $n$ sufficiently large, it holds that for any $W_{Z|XY}$ 
and any $P\in \calP_n(\calX)$,
\begin{flalign}
e_n^q(R,P,W_{Y|X})
&\leq 
\overline{E}^{q}(R-\epsilon_n,P,W_{YZ|X})+\epsilon.
\label{eq: dfhvdiudsvhidfiludffdhliufihi}
\end{flalign}

\end{theorem}

Note that $W_{YZ|X}=W_{Y|X}\times W_{Z|XY}$, and that any choice of $W_{Z|XY}$ in (\ref{eq: dfhvdiudsvhidfiludffdhliufihi}) is valid, 
and can be optimized as a function of $(P,R,q,W_{Y|X},\epsilon,n)$. 
In Theorem \ref{th: strict inprovement thm} to follow, we present a certain choice of $W_{Z|XY}$ which results in a strictly improved bound compared to several known ones. 
The proof of Theorem \ref{th: Main Theorem } can be found in Section \ref{sc: Proof of Exponent Theorem }. 

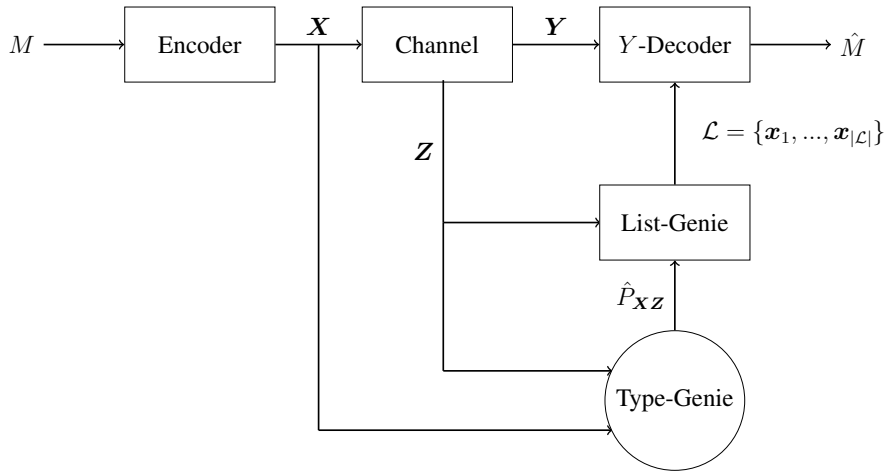
\begin{figure}[H]
	\resizebox{0.7\columnwidth}{!}{\input{Figure_BC_genie_2rec.tex}}
	\caption{The genie-aided-genie proof technique.}.
	\label{Figure_BC_genie}
\end{figure}

\noindent{\bf Proof Outline:} As mentioned above, the bound relies on the approach of multicast transmission with an auxiliary receiver.  
 The idea behind this proof technique is very simple (see Fig.\ \ref{Figure_BC_genie}). 
We extend the single-user channel from $X$ to $Y$ to have an additional output $Z$. 
The $Z$-receiver, serves as a genie to the $Y$-receiver, and provides it with the list 
 of all the codewords, which lie in the same conditional type-class given the received signal $\bZ$ as that of the true transmitted one, including the latter. 
To accomplish this, another genie informs the $Z$-receiver of the actual joint distribution of $\bZ$ and the true codeword $\bX$. Thus, the $Z$-receiver can be viewed as a ``genie-aided-genie".

The $Y$-receiver needs only to search within the narrowed list for the codeword that maximizes the decoding metric (either maximum likelihood or mismatched) rather than within the entire codebook, and therefore, analyzing this setup yields an upper bound on the exponent of the error probability.

We further show that provided that the list size is large enough, most of the list can be partitioned into disjoint subsets (sub-lists) of large size, where within each sub-list, the pairwise joint distribution given $\bz$ is approximately symmetric and approximately the same for all pairs. This extends a result of \cite{BondaschiDalai2022}, which can be viewed as the existence of such single sub-list for the case where $\bz$ is null.

This above partitioning of the list enables to show that if one considers channels with $\bz$ output for which with high probability the list size is large enough, then the average error probability conditioned on $\bz$ is essentially lower bounded by the average pairwise error probability of codewords within a sub-list. The latter can be lower bounded using Plotkin's counting trick as in  \cite{SHANNONGallagerBerlekamp1967522} and the above mentioned Blinkovsky's idea \cite{Blinovsky2002} of using Ramsey-theoretic result by Koml\'{o}s \cite{komlos1990strange} similar to \cite{BondaschiDalai2022,BondaschiGuilleniFabregasDalai-IT2021}.

\subsection{A Dual Form of the Bound}\label{sc: Dual Form of the Bound}

Next, we present a semi-dual form of the bound $\overline{E}^{q}(R,P,W_{YZ|X})$, whose proof appears in Appendix \ref{ap: DUAL EXP APPENDIX}.  
\begin{proposition}\label{pr: dual}
\begin{flalign}
&\overline{E}^{q}(R,P,W_{YZ|X})= \min_{\substack{P_{XZ}:P_X=P,\\ I(X;Z)\leq R }}
D(P_{Z|X}\|W_{Z|X}|P)+ \max_{P_{U|XZ}} \eta_q(P_{XZU},W_{Y|XZ})
\label{eq: ofauhvoidfhv}
\end{flalign}
where 
\begin{flalign}
& \eta_q(P_{XZU},W_{Y|XZ})\triangleq \nonumber\\
& \sup_{s\geq 0}-\sum_{u,\widetilde{x},x,z}P_{XZU}(x,z,u)P_{X|ZU}(\widetilde{x}|u,z)\log \sum_yW(y|x,z)e^{s[q(\widetilde{x},y)-q(x,y)]}.\label{eq: aiugvliufdglivugdfuiv}
\end{flalign}
\end{proposition}

Consider the function indexed by $(z,s)$, which can be thought of as a distance between two input symbols:
\begin{flalign}
d_{z,s,q,W_{Y|XZ}}(x,\widetilde{x})&= -\frac{1}{2}\log \left[\left(\sum_yW(y|x,z)e^{s[q(\widetilde{x},y)-q(x,y)]}\right)
\left(\sum_{y'}W(y'|\widetilde{x},z)e^{s[q(x,y')-q(\widetilde{x},y')]}\right)\right].\label{eq: d z s q dfn}
\end{flalign}

Note that $\eta_q(P_{XZU},W_{Y|XZ})$ can be expressed as
\begin{flalign}
& \eta_q(P_{XZU},W_{Y|XZ})=\sup_{s\geq 0}\sum_{u,\widetilde{x},x,z}P_{XZU}(x,z,u)P_{X|ZU}(\widetilde{x}|u,z) d_{z,s,q,W_{Y|XZ}}(x,\widetilde{x}).\label{eq: aiugvliufdglivugdfuiv}
\end{flalign}

\subsection{A Comparison to Previous Results}\label{sc: Comparison to Previous Results}

The following theorem shows that our bound recovers some known results, and can be strictly tighter than the straight-line bound (\ref{eq: SLLSP}). 

Define the following quantity that does not depend on $R$:
\begin{flalign}
\overline{E}_0^q(P,W) &\triangleq \max_{\substack{V_{\widetilde{X}UX}:\\
V_{X}=P\\
V_{XU}= V_{\widetilde{X}U}\\
\widetilde{X}-U-X
}}
\min_{ \substack{\widetilde{V}_{X\widetilde{X}Y}:\\ \widetilde{V}_{X\widetilde{X}}=V_{X\widetilde{X}}\\
\EE_{\widetilde{V}}q(\widetilde{X},Y)\geq \EE_{\widetilde{V}} q(X,Y)
} }D(\widetilde{V}_{Y|X\widetilde{X}}\|W_{Y|X}|V_{X\widetilde{X}})\label{eq: aoufhv;ufdhv}
\end{flalign}
Further, define the two-output (broadcast) channel 
\begin{flalign}
W_{YZ|X}^{(\alpha)}= W_{Y|X}\times [\alpha\cdot\indicator_{\{Z=Y\}}+(1-\alpha)\cdot Q_Z],\label{eq: W chann alpha dfn}
\end{flalign}
where $W_{Y|X}=W$, $\calZ=\calY$, 
the notation $\indicator_{\{\calA\}}$ stands for the indicator function of an event $\calA$, and\footnote{Note that a similar result holds for other choices of $Q_Z$ such as $Q_Z(y)=\sum_xP(x)W(y|x)$
} $Q_Z(y)=\frac{1}{|\calY|}$ $\forall y\in\calY$.

Recall the definitions of the previous bounds in (\ref{eq: E_esp})-(\ref{eq: SLLSP}). 
\begin{theorem}\label{th: strict inprovement thm}
For all $\alpha\in[0,1]$, 
\begin{flalign}
\overline{E}^q(\alpha R,P,W_{YZ|X}^{(\alpha)})&\leq (1-\alpha)\overline{E}_0^q(P,W)+\alpha E_{sp}(R,P,W)
,\label{eq: sl proof}
\end{flalign}
with a strict inequality when $\alpha\in (0,1)$. 

Consequently, if $(q,W)$ is a balanced pair then 
\begin{flalign}
&E^q(0^+,W)=\max_{P\in\calP(\calX)}\overline{E}_0^q(P,W)=E_{ex}^q(0^+,W),
 \label{eq: fiuvb}
\\
 & \max_{P\in\calP(\calX)}
\left.\overline{E}^{q}(R,P,W_{YZ|X})\right|_{q=\log W,\; W_{Z|XY}=\indicator_{\{Z=Y\}}}\leq E_{sp}(R,W),\label{eq: audfgviufdg}
\end{flalign}
\end{theorem}
The proof of Theorem \ref{th: strict inprovement thm} can be found in Appendix \ref{sc: strict inprovement thm}, and as mentioned above, the equality $E^q(0^+,W)=E_{ex}^q(0^+,W)$ was proved in \cite{BondaschiGuilleniFabregasDalai-IT2021}. 
Note that the choice of $W_{YZ|X}^{(\alpha)}$ defined in (\ref{eq: W chann alpha dfn}) is not necessarily optimal, but rather serves our purpose of proving improvement of the bound compared to known results. 
\begin{corollary}
If the supremum in $\sup_{P\in\calP(\calX)} \overline{E}(R,P, W_{YZ|X}^{(\alpha_{\mbox{\tiny{R}}} )})$ is attained, then 
the following strict inequality holds
\begin{flalign}
 \max_{P\in\calP(\calX)} \overline{E}(R,P, W_{YZ|X}^{(\alpha_{\mbox{\tiny{R}}} )})&< E_{sl-sp}(R,W),\label{eq: aduvg;diufgv}
\end{flalign}
where $\alpha_{\mbox{\tiny{R}}} =R/R_W^*$. 
\end{corollary}
\begin{proof}
Since according to Theorem \ref{th: strict inprovement thm}, (\ref{eq: sl proof}) holds with strict inequality for any $P$ and $\alpha\in(0,1)$, 
substituting $q=q_{\mbox{\tiny{ML}}} =\log W$, and taking the maximum of $P$ over the two sides of the inequality yields
\begin{flalign}
&\overline{E}(\alpha R,W_{YZ|X}^{(\alpha)})<\alpha \cdot E_{sp}(R,W)+(1-\alpha) \cdot E_{ex}(0^+,W).
\end{flalign}
Taking 
$R=R_W^*$, this implies that our bound is strictly lower than the linear curve of the straight line bound; that is, (\ref{eq: aduvg;diufgv}) holds. 
\end{proof}
It is conjectured that in most cases the above mentioned supremum is attained, and thus the strict inequality (\ref{eq: aduvg;diufgv}) usually does hold. Proving this involves characterizing cases where $\overline{E}(R,P, W_{YZ|X}^{(\alpha_{\mbox{\tiny{R}}} )})$ is continuous in $P\in\calP(\calX)$. Moreover, there are clear cases for which one can determine the optimal $P$; e.g., the binary symmetric channel (BSC) case with ML decoding where due to symmetry considerations, at least for even $n$ one can prove that the optimal $P$ is equal to $[\frac{1}{2},\frac{1}{2}]$ (and for odd $n$'s one can use an identical $n$'s symbol for all codewords). In this case, Theorem \ref{th: strict inprovement thm} implies the strict inequality
\begin{flalign}
E(R,W_{Y|X})&
\leq\left. \overline{E}(R,P, W_{YZ|X}^{(\alpha_{\mbox{\tiny{R}}} )})\right|_{P=[\frac{1}{2},\frac{1}{2}]}<\alpha_{\mbox{\tiny{R}}}\cdot E_{sp}(R,W)+(1-\alpha_{\mbox{\tiny{R}}}) \cdot E_{ex}(0^+,W).\label{eq: binary first}
\end{flalign}

Next, we show that our bound is at least as tight as our previous best known upper bound (see (\ref{eq: sp 1})).
\begin{proposition}\label{pr: comarison to previous sp}
\begin{flalign}
\inf_{W_{Z|XY}}\overline{E}^q(R,P,W_{YZ|X})&\leq  E_{sp}^q(R,P,W).
\end{flalign}
\end{proposition}
Proposition \ref{pr: comarison to previous sp} is proved in Appendix \ref{sc: comarison to previous sp}.


In the next proposition we present a looser bound that takes a simpler form for both the matched and mismatched metric cases. Denote
\begin{flalign}
E_{B}^q(R,P,W)&\triangleq 
\min_{\substack{P_{XZ}\in\calP(\calX\times\calZ):\\
P_X=P,\\  I(X;Z)\leq R }}
 \max_{P_{U|XZ}}
 \sum_{u,\widetilde{x},x,z}P_{XZU}(x,z,u)P_{X|UZ}(\widetilde{x}|u,z)\eta_q(P_{XZU},W_{Y|X})\label{eq: aiduvgilfudgv}\\
 E_{B}^{q_{\mbox{\tiny{ML}}}}(R,P,W)&\triangleq 
\min_{\substack{P_{XZ}\in\calP(\calX\times\calZ):\\
P_X=P,\\  I(X;Z)\leq R }}
 \max_{P_{U|XZ}}
 \sum_{u,\widetilde{x},x,z}P_{XZU}(x,z,u)P_{X|UZ}(\widetilde{x}|u,z)\log \frac{1}{\sum_y\sqrt{W(y|x)W(y|\widetilde{x})} }
 ,
\end{flalign}
where the difference between (\ref{eq: ofauhvoidfhv}) and (\ref{eq: aiduvgilfudgv}), is that $\eta_q(P_{XZU},W_{Y|X})$ replaces $\eta_q(P_{XZU},W_{Y|XZ})$; i.e., in (\ref{eq: ofauhvoidfhv}) we have $W_{Y|XZ}$ instead of $W_{Y|X}$. 
We have the following result:
\begin{proposition}\label{eq: EB inequality}
The following inequality holds
\begin{flalign}
\inf_{W_{Z|XY}}\overline{E}^q(R,P,W_{YZ|X})\leq 
E_{B}^q(R,P,W),
\end{flalign}
and in particular 
\begin{flalign}
\inf_{W_{Z|XY}}\overline{E}(R,P,W_{YZ|X})\leq  E_{B}^{q_{\mbox{\tiny{ML}}}}(R,P,W),
\end{flalign}
with cases of strict inequality. 
\end{proposition}
The proof appears in Appendix \ref{sc: EB Inequaliity}. 
In Appendix \ref{ap: CORRECTING BLAHUT} we outline an alternative proof for the looser bound $E(R,P,W)\leq  E_{B}^{q_{\mbox{\tiny{ML}}}}(R,P,W)$ that is based mostly on the derivation of the bound (\ref{eq: 16sfhivdhv}) of \cite{BondaschiDalai2022}. 
It is interesting to note that in this alternative proof of the looser bound, the output signal $\bZ$ is substituted by the deterministic sequence whose existence is guaranteed by a covering argument (see \cite[Lemma 5]{BondaschiDalai2022}, and \cite[Lemma 1]{BlahutComposition}).

\subsection{A Looser Bound and an Example: the BSC with ML Decoder}\label{sc: Introduction of a Looser Bound and an Example: the BSC with ML Decoder}

It is desirable to evaluate the bound $\overline{E}^{q}(R,P,W_{YZ|X})$ or simplify it, and in particular, to tackle the maximization over $P_{U|XZ}$. 
The first obvious thing to do is upper bound this maximization by a maximization over a certain symmetric set of marginal distributions $P_{\widetilde{X}ZX}$.
Let
\begin{flalign}\label{eq: sym dfn set}
\calP_{sym}(\calX^2\times \calZ)&\triangleq \{P_{XZ\widetilde{X}}\in \calP(\calX^2\times \calZ):\; \forall (x,\widetilde{x}, z),\; P_{XZ\widetilde{X}}(x,z,\widetilde{x})=P_{\widetilde{X}XZ}(x,z,\widetilde{x})\},
\end{flalign}
and define
\begin{flalign}
& \widetilde{\eta}_q(P_{XZ\widetilde{X}},W_{Y|XZ})=\sup_{s\geq 0}\sum_{\widetilde{x},x,z}P_{XZ\widetilde{X}}(x,z,\widetilde{x}) d_{z,s,q,W_{Y|XZ}}(x,\widetilde{x}).\label{eq: aiugvgdfuiv}
\end{flalign}

We have the following upper bound, it is proved in Appendix \ref{cs: U exepmt corllary}.
\begin{corollary}\label{eq: U exepmt corllary}
\begin{flalign}
&\overline{E}^{q}(R,P,W_{YZ|X})\leq\overline{E}_{sym}^{q}(R,P,W_{YZ|X})\triangleq \nonumber\\
&\min_{\substack{P_{XZ}:\\
P_X=P,\\  I(X;Z)\leq R }}
\max_{\substack{P_{\widetilde{X}|XZ}:\\
P_{XZ\widetilde{X}}\in \calP_{sym}(\calX^2\times \calZ)\\
\forall (x,z),\; P_{\widetilde{X}X|Z}(x,x|z)\geq P^2_{X|Z}(x|z)
}}
\min_{ \substack{ P_{Y|XZ\widetilde{X}}:\\ 
\EE q(\widetilde{X},Y)\geq \EE q(X,Y)
} } 
D(P_{YZ|X}\|W_{YZ|X}|P)+I(\widetilde{X};Y|X,Z)\label{eq: aifudgviufdgvd1}\\
&= \min_{\substack{P_{XZ}:\\
P_X=P,\\  I(X;Z)\leq R }}D(P_{Z|X}\|W_{Z|X}|P)+
\max_{\substack{P_{\widetilde{X}|XZ}:\\ 
P_{XZ\widetilde{X}}\in \calP_{sym}(\calX^2\times \calZ)\\
\forall (x,z),\; P_{\widetilde{X}X|Z}(x,x|z)\geq P^2_{X|Z}(x|z)
}}
\widetilde{\eta}_q(P_{XZ\widetilde{X}},W_{Y|XZ}).\label{eq: aifudgviufdgvd}
\end{flalign}
\end{corollary}
In the case of the BSC, we already showed that the strict inequality (\ref{eq: binary first}) holds. 
We next show that the bound $\overline{E}_{sym}^{q}(R,P,W_{YZ|X})$ of Corollary \ref{eq: U exepmt corllary} takes the following simple form:
\begin{flalign}
\min_{\substack{P_{XZ}:P_X=P,\\ I(X;Z)\leq R }}
D(P_{Z|X}\|W_{Z|X}|P)+ \max_{\substack{P_{\widetilde{X}|XZ}:\\ 
P_{XZ\widetilde{X}}\in \calP_{sym}(\calX^2\times \calZ)\\
\forall (x,z),\; P_{\widetilde{X}X|Z}(x,x|z)\geq P^2_{X|Z}(x|z)
}}
\sup_{s\geq 0}\sum_{z,x,\widetilde{x}}P_Z(z)\frac{\Pr(X\neq \widetilde{X}|z)}{2}\cdot d_{z,s,q,W_{Y|XZ}}(x,\widetilde{x})
\end{flalign}
Taking $P=[1/2,1/2]$, restricting to $P_{Z|X}(0|z)= P_{Z|X}(1|z)\triangleq \gamma$, and using the constraint $P_{\widetilde{X}X|Z}(x,x|z)\geq P^2_{X|Z}(x|z)$, we have $\Pr(X\neq \widetilde{X}|z)\leq 1-P^2_{X|Z}(0|z)-P^2_{X|Z}(1|z)=2\gamma(1-\gamma)$, and this yields the bound
\begin{flalign}
&\min_{\gamma:\; 1-h_2(\gamma)\leq R }
D(P_{Z|X}\|W_{Z|X}|P)+  \sup_{s\geq 0}\sum_{x,\widetilde{x},z: d_{z,s,q,W_{Y|XZ}}(x,\widetilde{x})>0}P_Z(z)
\gamma(1-\gamma)\cdot d_{z,s,q,W_{Y|XZ}}(x,\widetilde{x})\nonumber\\
&=\min_{1-h_2(\gamma)\leq R }
\frac{1}{2}\bigg[\gamma\log\frac{\gamma}{W_{Z|X}(1|0)}+ \overline{\gamma}\log\frac{\overline{\gamma}}{W_{Z|X}(0|0)}+
\gamma\log\frac{\gamma}{W_{Z|X}(0|1)}+ \overline{\gamma}\log\frac{\overline{\gamma}}{W_{Z|X}(1|1)}\bigg]\nonumber\\
&+  \gamma(1-\gamma)\frac{1}{2}\sup_{s\geq 0}\bigg[\sum_{z:\;d_{z,s,q,W_{Y|XZ}}(0,1)>0}
 d_{z,s,q,W_{Y|XZ}}(0,1)+\sum_{z:\;d_{z,s,q,W_{Y|XZ}}(1,0)>0}
d_{z,s,q,W_{Y|XZ}}(1,0)\bigg] ,\label{eq: BSC found bound}
\end{flalign}
where $h_2(x)=-x\log(x)-(1-x)\log(1-x)$ is the binary entropy function, and $d_{z,s,q,W_{Y|XZ}}(x,\widetilde{x})$ is defined in (\ref{eq: d z s q dfn}).

\begin{figure}\label{fig: BSC ML}
\centering{\includegraphics[width=0.7\linewidth]{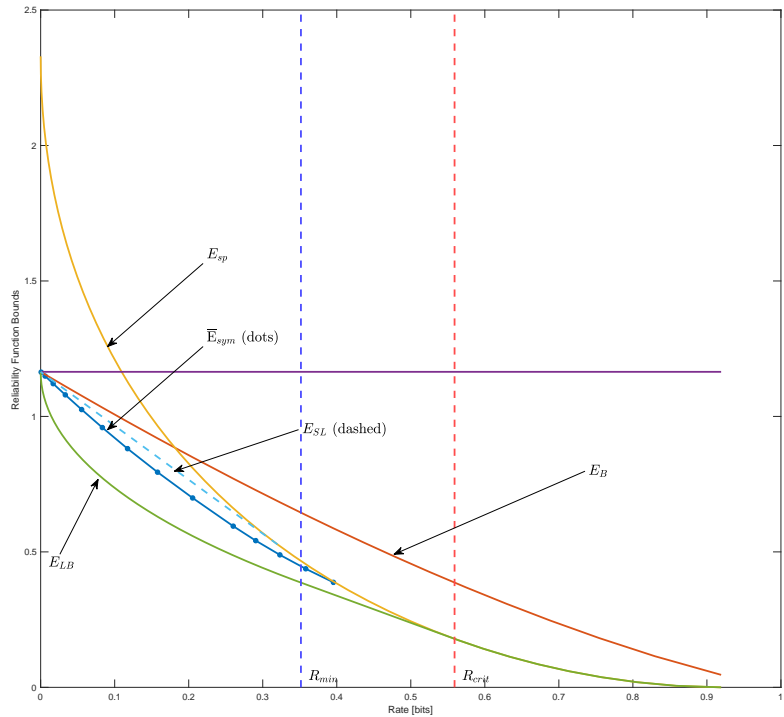}}
\caption{Bounds on the reliability function of the BSC with crossover probability $p=0.1$. 
$E_{LB}$ (\ref{eq: ELB}), $\overline{E}_{sym}$ (\ref{eq: BSC found bound}), $E_{sp}$ (\ref{eq: E_esp}),  
} 
\end{figure}

A numerical calculation of the bound (\ref{eq: BSC found bound}) for $|\calZ|=2$ is depicted in Fig.\ \ref{fig: BSC ML} which is calculated using a grid search for the optimal choice of $W_{Z|XY}$, however, since any value of $W_{Z|XY}$ yields a valid upper bound, this implies that the actual bound (\ref{eq: BSC found bound}) may be tighter than the one depicted. 
Moreover, the bound $\inf_{W_{Z|XY}}\overline{E}^{q}(R,P,W_{YZ|X})$ may be even tighter, and in addition $|\calZ|$ can be taken larger than $2$. 
For the sake of comparison, the upper bounds $E_{sp}(R)$, and the amended Blahut bound $E_B(R)$ are also depicted, as well as the straight line bound, and the lower bound:
\begin{flalign}\label{eq: ELB}
E_{LB}(R)=\left\{\begin{array}{ll} 
-\delta_{GV}(R)\cdot \log\left(2\sqrt{p(1-p)}\right)
& R\leq R_{min}\\
1-\log\left(1+2\sqrt{p(1-p)}\right)-R & R\in [R_{min},R_{crit}]\\
E_{sp}(R) &R\geq R_{crit}
\end{array}\right.
\end{flalign}
where $\delta_{GV}(R)$ is the solution to the equation $1-h_2(\delta)=R$, and $R_{min}=1-h_2\left(\frac{2\sqrt{p(1-p)}}{1+2\sqrt{p(1-p)}}\right)$. Exploiting symmetries as well as other special properties, tighter bounds compared to the general DMC case have been derived for the BSC, which are not depicted in this figure (see \cite{burnashev2016reliability}), and seem tighter compared to our numerical calculation.  
In order to compare $\inf_{W_{Z|XY}}\overline{E}^{q}(R,P,W_{YZ|X})$ with larger $|\calZ|$ (rather than the looser version (\ref{eq: BSC found bound}) with $|\calZ|=1$) applied to the BSC case to the bound of \cite{burnashev2016reliability}, further study is needed. 
Nevertheless, the BSC serves as an example that shows that our bound strictly improves on the straight line bound as well as the Blahut amended bound and the sphere packing bound.

\subsection{An Approximation - a Function that Lies Between the Upper Bound and the Lower Bound}\label{sc: Approximating Function - a Function that Lies Between our Upper Bound and the Lower Bound}

 Recall the definition of $E^q_{\mbox{\tiny{CK}}}(R,P,W)$ in (\ref{eq: E_ex dfn C and K 22 22 udgfugeug}). 
 The following result is a corollary of Theorem \ref{th: Main Theorem } and Proposition \ref{pr: dual}. It introduces the function $E_{orth}^{q}(R,P,W_{Y|X})$ which lies between the lower and upper bounds on $E^q(R,P,W)$. 
\begin{corollary}\label{cr: quadratic}
\begin{flalign}
\inf_{W_{Z|XY}}\overline{E}^{q}(R,P,W_{YZ|X})\geq E_{orth}^{q}(R,P,W_{Y|X})\geq E^q_{\mbox{\tiny{CK}}}(R,P,W) ,\label{eq: aiuviufhdiuvhdfiuhviudf}
\end{flalign}
where
\begin{flalign}
E_{orth}^{q}(R,P,W_{Y|X})
&=\min_{\substack{P_{\widetilde{X}XYZ}:\\
P_X=P,\;  I(X;Z)\leq R \\
\EE q(\widetilde{X},Y)\geq \EE q(X,Y),\\P_{XZ}=P_{\widetilde{X}Z},\; X-Z-\widetilde{X}
} } 
D(P_{Y|X}\|W_{Y|X}|P)+I(\widetilde{X};Y|X,Z).\label{eq: a'voiha'odisfjvp'ojdf'povopdfkv}
\end{flalign}
\end{corollary}
Corollary \ref{cr: quadratic} is proved in Appendix \ref{sc: Proof of Corollary Blahut}. 
When the gap between the lower and upper bounds is small, $E_{orth}^{q}(R,P,W_{YZ|X})$ can be regarded as a relatively simple expression for an approximation of the reliability function. 

\section{Proof of Theorem \ref{th: Main Theorem }}\label{sc: Proof of Exponent Theorem }

Let $n$ be the block length, fix $P_n\in\calP_n(\calX)$, and let $\calC_n=\{\overline{\bx}_i\}_{i=1}^{\mathbb{M}_n}\subseteq \calT_n(P_n)$ be a $P_n$-constant composition codebook of rate $R$ for the channel $W_{Y|X}$. 
Let $\calZ$ be such that $|\calZ|<\infty$, let $\epsilon>0$, and let a DMC $W_{YZ|X}$ be given, for which $W=W_{Y|X}$ and $q$ are balanced.
We choose small enough ${\delta}>0$ 
to be specified later.

For technical reasons, we assume that the transmission takes place over the two-outputs discrete memoryless channel ${W}^{{\delta}}_{YZ|X}$ which slightly differs from $W_{YZ|X}$, where:
\begin{flalign}\label{eq: ahfuv W star}
\forall (x,y,z),\; {W}^{{\delta}}_{YZ|X}(y,z|x)&= W_{Y|X}(y|x)\cdot\left[(1-{\delta}) \cdot W_{Z|XY}(z|x,y)+{\delta}\cdot \frac{1}{|\calZ|}\right].
\end{flalign}
Note that 
$\forall z\in\calZ,\; {W}^{{\delta}}_{Z|X}(z|x)>\frac{{\delta}}{|\calZ|} $, and further
\begin{flalign}\label{eq: ahfuv W starert7eft}
\forall (x,y,z),\; {W}^{{\delta}}_{Y|XZ}(y|x,z)&= \frac{W(y|x){W}^{{\delta}}(z|x,y)}{{W}^{{\delta}}(z|x)}\geq W_{Y|X}(y|x)\cdot \frac{{\delta}}{|\calZ|}. 
\end{flalign}
It is easy to verify that this implies that also the marginal channel ${W}^{{\delta}}_{Y|XZ}$ from $\calX\times \calZ$ to $\calY$ along with the metric $q((x,z),y)=q(x,y)$ are balanced (see (\ref{eq: balanced C 0 conditions})). 
%
%
%
%
Note also that from (\ref{eq: ahfuv W starert7eft}) it follows that
\begin{flalign}\label{eq: ahfuadvdvdsv W starert7eft}
\forall (x,y,z),\; 
W(y|x)&=\frac{{W}^{{\delta}}(z|x)}{{W}^{{\delta}}(z|x,y)}
{W}^{{\delta}}_{Y|XZ}(y|x,z)\geq {W}^{{\delta}}(y|x,z)\cdot \frac{{\delta}}{|\calZ|}.
\end{flalign}

Recall the definition of empirical distributions and types (which appears after (\ref{eq: afuddiadvadsvudgfuvf})). 
The conditional type-class of $\bz$ given $\bx$ is the set of sequences $\tilde{\bz}$'s such that $\widehat{P}_{\bx\tilde{\bz}}=\widehat{P}_{\bx\bz}$, which is denoted $\calT_n(\widehat{P}_{\bz|\bx} |\bx)$.

Given the channel input $\bX\in \calT_n(P_n)$, which is uniformly distributed over $\calC_n$, and the joint empirical distribution of $(\bX,\bZ)$, $\widehat{P}_{\bX\bZ}=\widehat{P}_{XZ}$, we clearly have that $\bZ$ is distributed uniformly over $\calT_n(\widehat{P}_{Z|X}|\bX)$; i.e.,
\begin{flalign}\label{eq: mu channel}
\Pr(\bZ=\bz|\bX=\bx,\widehat{P}_{\bX\bZ}=\widehat{P}_{XZ})&=\frac{\indicator_{\{\bz\in \calT_n(\widehat{P}_{Z|X}|\bx)\}}}{|\calT_n(\widehat{P}_{Z|X}|\bx)|}.
\end{flalign}

Let
\begin{flalign}\label{eq: List dfn with type}
\calL(\bz,\widehat{P}_{\bx\bz})&\triangleq \calC_n\cap\calT_n(\widehat{P}_{\bx|\bz}|\bz)\nonumber\\
&\triangleq \big\{\bx_1(\bz,\widehat{P}_{\bx\bz}),\ldots,\bx_{|\calL(\bz,\widehat{P}_{\bx\bz})|}(\bz,\widehat{P}_{\bx\bz})\big\}
\end{flalign}
 be the list containing all the codewords, which lie in the same 
 conditional type-class given the received signal $\bZ$ as that of the transmitted one. 
Assume that the $Y$-decoder is informed of the list $\calL(\bZ,\widehat{P}_{XZ})=\{\bx_i(\bZ,\widehat{P}_{XZ})\}$ and employs the decoding rule
\begin{flalign}\label{eq: genie aided decoder}
\widehat{m}&=\argmax_{i:\; \overline{\bx}_i\in \calL(\bz,\widehat{P}_{\bx\bz})} q(\overline{\bx}_i,\by),
\end{flalign}
where ties are broken uniformly between the maximizers.

As in \cite{SomekhBaruchArxiv_16March2022}, it is easily verified that for any possible channel output $\bz$ such that $\widehat{P}_{\bz}=\widehat{P}_Z$, it holds that 
$\{\bx_i(\bz,\widehat{P}_{XZ})\}$ are equiprobable given $\{\bZ=\bz,\widehat{P}_{\bX\bZ}=\widehat{P}_{XZ}\}$; that is, 
\begin{flalign}\label{eq: idfuhvilufg}
 P(\bX=\bx_i(\bz,\widehat{P}_{XZ})|\bZ=\bz,\widehat{P}_{\bX\bZ}=\widehat{P}_{XZ})=\frac{1}{|\calL(\bz,\widehat{P}_{XZ})|}.
\end{flalign}
For the sake of completeness, we also include the proof of (\ref{eq: idfuhvilufg}) in Appendix \ref{ap: uniformity}.

Next, let 
\begin{flalign}
\calE_{ij}\triangleq \calE_{ij}(\bz,\widehat{P}_{XZ})\triangleq \{\by:\; q(\bx_j(\bz,\widehat{P}_{XZ}),\by)\geq q(\bx_i(\bz,\widehat{P}_{XZ}),\by)\},
\end{flalign}
and adopt the shorthand notation
\begin{flalign}
\bx_i\triangleq \bx_i(\bz,\widehat{P}_{XZ}),\; \calL\triangleq \calL(\bz,\widehat{P}_{XZ}). 
\end{flalign}
Since $\Pr(error|\bx_i,\bz)= \Pr(\cup_{j\neq i}\calE_{ij}|\bx_i,\bz)$, 
we have the following lower bound\footnote{In \cite{SomekhBaruchArxiv_16March2022}, a similar (but looser) initial bound was given with $\max_{j\in\calL:\; j\neq i}
\Pr(\calE_{ij}|\bx_i,\bz)$ which appear is (\ref{eq: STAM}) replaced by $\frac{1}{|\calL|-1}\sum_{j\in\calL:\; j\neq i}
\Pr(\calE_{ij}|\bx_i,\bz)$.} on the average error probability in $q$-mismatched decoding at the $Y$-receiver given that $\bZ=\bz$ and $\widehat{P}_{\bX\bZ}=\widehat{P}_{XZ}$, 
\begin{flalign}\label{eq: STAM}
\Pr(error|\bz,\widehat{P}_{\bX\bZ}=\widehat{P}_{XZ})&\geq\left\{\begin{array}{ll} \frac{1}{|\calL|}\sum_{i\in\calL} \max_{j\in\calL:\; j\neq i}
\Pr(\calE_{ij}|\bx_i,\bz)&|\calL|\geq 2\\
0&|\calL|=1\end{array}\right. .
\end{flalign}

Next, let $\hat{\epsilon}_n$ be a vanishing sequence which satisfies $\lim_{n\rightarrow \infty}n\hat{\epsilon}_n=\infty$ (i.e., $\hat{\epsilon}_n>> \frac{1}{n}$).
We confine attention to such $\bz\in\calT_n(\widehat{P}_Z)$ which satisfy $|\calL|\geq e^{n\hat{\epsilon}_n}$.

Now, let $\calG$ be a subset of $\calL$ which is partitioned into $K$ disjoint subsets $\{\calG_k\}_{k=1}^K $; i.e., 
\begin{flalign}
\calG&=\cup_{k=1}^K \calG_k\subseteq \calL,\\ &\calG_k\cap\calG_\ell=\emptyset,\; \forall k\neq \ell. 
\end{flalign}
Therefore, in this case we have 
\begin{flalign}
&\frac{1}{|\calL|}\sum_{i\in\calL} \max_{j\in\calL:\; j\neq i}
\Pr(\calE_{ij}|\bx_i,\bz)\nonumber\\
&= \frac{1}{|\calL|}
\sum_{k=1}^N
\sum_{i\in\calG_k} 
\max_{j\in\calL:\; j\neq i}\Pr(\calE_{ij}|\bx_i,\bz)\\
&\geq \frac{1}{|\calL|}
\sum_{k=1}^N
\sum_{i\in\calG_k} 
\frac{1}{|\calG_k|-1} \sum_{j\in\calG_k:\; j\neq i}\Pr(\calE_{ij}|\bx_i,\bz)\\
&= \frac{|\calG|}{|\calL|}
\sum_{k=1}^N\frac{|\calG_k|}{|\calG|} 
\sum_{i,j\in\calG_k:\;  j\neq i} 
\frac{1}{|\calG_k|(|\calG_k|-1)} \Pr(\calE_{ij}|\bx_i,\bz)\label{eq: STAMaekurghviuh}
\end{flalign}

In what follows we present a few results and definitions that will be used to determine the sets $\{\calG_k\}$ that are used in our proof. 

\begin{definition}\label{df: aidjosj}
Let $\bz\in\calZ^n$ and $\widehat{P}_{X|Z}$ be given. A codebook $\{\bx_i\}\subseteq\calT_n(\widehat{P}_{X|Z}|\bz)$ of size $M'$ which satisfies \begin{flalign}\label{eq: idivugfiv}
|\hat{P}_{\bx_m\bz\bx_j}(x,z,x')-\hat{P}_{\bx_\ell\bz\bx_k}(x,z,x')|\leq 6/\sqrt{M'}+ 2\sqrt{2}/\sqrt{t} +3/t
\end{flalign}\noindent for all $i\neq j$ and
$\ell\neq k$ (not necessarily different from $m$ and $j$) and any $(x, x')\in\calX^2$ and $z\in\calZ$, is said to be \underline{$(t,\widehat{P}_{XZ})$-quasi-symmetric w.r.t.\ $\bz$}.
\end{definition}

The following lemma generalizes the result of \cite[Theorem 3]{BondaschiGuilleniFabregasDalai-IT2021} which is based on a result by Koml\'{o}s \cite{komlos1990strange}, to the case of a non null sequence $\bz$.
\begin{lemma}\label{lm: Komlos Lemma} 
For any positive integers $t$ and $M'$, there exists a
positive integer $M_0(M' , t)$ such that for any $\bz\in\calZ^n$ and $\widehat{P}_{X|Z}$, any code $\calC_n\subseteq\calT_n(\widehat{P}_{X|Z}|\bz)$ of $ |\calC_n| > M_0(M' , t)$ codewords, contains a sub-code $\calC_n'\subseteq \calC_n$ with $M'$ codewords which is $(t,\widehat{P}_{XZ})$-quasi-symmetric w.r.t.\ $\bz$.
\end{lemma}\label{lm: Komlos Lemma generalized}
\begin{proof}
Consider the codebook $\calC_n=\{\bv_1,...,\bv_M$\} with $\bv_m=(\bx_m,\bz)$, then apply \cite[Theorem 3]{BondaschiGuilleniFabregasDalai-IT2021} to the codebook $\calC_n$. 
\end{proof}
The following lemma is an outcome of Lemma \ref{lm: Komlos Lemma}.
\begin{lemma}\label{lm: Komlos Lemma generalized second lemma} ({\bf Code Partitioning to quasi-symmetric sub-codes})
For any positive integers $t$, and $M'$ there exists a
positive integer $M_0(M' , t)$ such that for any integer $J\geq 2$, $\bz\in\calZ^n$ and $\widehat{P}_{X|Z}$, any code $\calC_n\subseteq\calT_n(\widehat{P}_{X|Z}|\bz)$ of $ |\calC_n| > J\cdot \max\left\{ M_0(M' , t),M'\right\}$ codewords, contains $N=\lfloor \frac{|\calC_n|- \max\left\{M_0(M' , t),M'\right\}}{M'}\rfloor $ disjoint sub-codes $\calC_{n,i}'\subseteq \calC_n$, each with $M'$ codewords, and each is $(t,\widehat{P}_{XZ})$-quasi-symmetric w.r.t.\ $\bz$, and consequently
\begin{flalign}
\left|\cup_{i=1}^N \calC_{n,i}'\right|\geq (1-\frac{2}{J})|\calC_n|.
\end{flalign}
\end{lemma}
\begin{proof}
Fix $(t,M')$, and the resulting $M_0(M',t)$ defined in Lemma \ref{lm: Komlos Lemma}, fix also $J\geq 2$, and let a codebook $\calC_n$ such that $ |\calC_n| > J\cdot \max\left\{M_0(M' , t),M'\right\}$ be given. 
Extract from $\calC_n$ a sub-code which is $(t,\widehat{P}_{XZ})$-quasi-symmetric w.r.t.\ $\bz$ of size $M'$, whose existence is ensured by Lemma \ref{lm: Komlos Lemma}. Denote the extracted sub-code by $\calC_{n,1}'$. 
Next, consider the collection of the remaining codewords of size $|\calC_n|-M'$. 
Since we have $|\calC_n|-M'> (J-1)\max\left\{M_0(M' , t),M'\right\}$, by Lemma \ref{lm: Komlos Lemma}
we can extract another sub-code which is $(t,\widehat{P}_{XZ})$-quasi-symmetric w.r.t.\ $\bz$ of size $M'$, and denote the extracted sub-code by $\calC_{n,2}'$. 
Continue recursively to extract a total of $N$ sub-codes, $\{\calC_{n,i}'\}_{i=1}^N$, which is possible as long as the condition $|\calC_n|-N\cdot M'\geq \max\left\{M_0(M' , t),M'\right\}$ holds; i.e., one can extract at least 
$N= \lfloor \frac{|\calC_n|- \max\left\{M_0(M' , t),M'\right\}}{M'}\rfloor$ such sub-codes. 
Note that
\begin{flalign}
\left|\cup_{i=1}^N \calC_{n,i}'\right|&= N\cdot M'\nonumber\\
&\geq \left(\frac{|\calC_n|- \max\left\{M_0(M' , t),M'\right\}}{M'}-1\right)\cdot  M'\\
&=
|\calC_n|- \max\left\{M_0(M' , t),M'\right\}-M'\\
&\geq 
|\calC_n|-\frac{1}{J}|\calC_n|-M'\\
&\geq (1-\frac{2}{J})|\calC_n|.
\end{flalign}
where the last two inequalities follows since $ |\calC_n| > J\cdot \max\left\{M_0(M' , t),M'\right\}$. 
\end{proof}


Now, fix sufficiently large integers $(t,M')$, and the resulting $M_0(M',t)$, take for example\footnote{Taking larger $J$ only yields a larger prefactor to the upper bound on the error probability, but taking $J=4$ suffices.} $J=4$, and let $n$ be such that 
\begin{flalign}
 e^{n\hat{\epsilon}_n}>4\cdot \max\left\{M_0(M',t),M'\right\}.\label{eq: avuh;difuhv;oihdf;oiv}
\end{flalign} 
Further, assume $|\calL|=|\calL(\bz,\hat{P}_{XZ})|\geq  e^{n\hat{\epsilon}_n}$, and apply Lemma \ref{lm: Komlos Lemma generalized second lemma} to the list $\calL(\bz,\hat{P}_{XZ})$ in the role of the codebook. 
Let $\calG_k=\calG_k(\bz,\hat{P}_{XZ})$, $k=1,...,N$ stand for the 
$(t,\widehat{P}_{XZ})$-quasi-symmetric w.r.t.\ $\bz$ 
sub-lists, whose existence is guaranteed by Lemma \ref{lm: Komlos Lemma generalized second lemma}; that is, 
\begin{flalign}
&\forall k, \; |\calG_k|= M'\\
&\left|\cup_{k=1}^N \calG_k\right|\geq \frac{1}{2}\cdot |\calL|.
\end{flalign}

Next, let $\calG=\cup_{k=1}^K \calG_k$, we have from (\ref{eq: STAMaekurghviuh}) 
\begin{flalign}
&\frac{1}{|\calL|}\sum_{i\in\calL} \max_{j\in\calL:\; j\neq i}
\Pr(\calE_{ij}|\bx_i,\bz)\nonumber\\
&\geq \frac{1}{2}
\sum_{k=1}^N\frac{|\calG_k|}{|\calG|} 
\sum_{i,j\in\calG_k:\;  j\neq i} 
\frac{1}{|\calG_k|(|\calG_k|-1)} \Pr(\calE_{ij}|\bx_i,\bz).\label{eq: alfgvilufdgv}\end{flalign}
Now, using standard method of types arguments, we have
\begin{flalign}
\Pr(\calE_{ij}|\bx_i,\bz)=& \sum_{\by:\; q(\bx_j,\by)\geq q(\bx_i,\by)}\left({W}^{{\delta}}_{Y|XZ}\right)^n(\by|\bx_i,\bz)\\
&\geq \frac{1}{(n+1)^{|\calX|^2|\calZ||\calY|}}\cdot 
\sum_{
\substack{V_{\widetilde{X}XZY}\in\calP_n(\calX^2\times\calZ\times\calY):\; \\
V_{XZ\widetilde{X}}=\widehat{P}_{\bx_i\bz\bx_j},\\ q(V_{\widetilde{X}Y})\geq q(V_{XY})}}e^{-n\cdot D(V_{Y|XZ\widetilde{X}}\|{W}^{{\delta}}_{Y|XZ}|\widehat{P}_{\bx_i\bz\bx_j})
}\\
&\geq \frac{1}{(n+1)^{|\calX|^2|\calZ||\calY|}}\cdot e^{-n\Omega^q_n(\widehat{P}_{\bx_i\bz\bx_j}, {W}^{{\delta}}_{Y|XZ})}.\label{eq: afhjviodfhvaifudhi}
\end{flalign}
where for an empirical distribution $P_{XZ\widetilde{X}}\in\calP_n(\calX^2\times\calZ)$ satisfying $P_{\widetilde{X}Z}=P_{XZ}$ we define
\begin{flalign}
&\Omega^q_n(P_{XZ\widetilde{X}}, {W}^{{\delta}}_{Y|XZ})\triangleq 
\min_{\substack{V_{\widetilde{X}XYZ}\in\calP_n(\calX^2\times\calZ\times\calY)\cap \calS_q(P_{XZ\widetilde{X}})}} D(V_{Y|XZ\widetilde{X}}\|{W}^{{\delta}}_{Y|XZ}|P_{XZ\widetilde{X}})\label{eq: Ialpha dfndfugilugi}\\
&\calS_q(P_{XZ\widetilde{X}})\triangleq \left\{V_{\widetilde{X}XYZ}\in\calP(\calX^2\times\calZ\times\calY):
V_{XZ\widetilde{X}}=P_{XZ\widetilde{X}},\; 
q(V_{\widetilde{X}Y})\geq q(V_{XY}) \right\}\label{eq: calSdfn}.
\end{flalign}
Denoting $A_n\triangleq \frac{1}{(n+1)^{|\calX|^2|\calZ||\calY|}}$, this yields, 
\begin{flalign}
&
\sum_{k=1}^N\frac{|\calG_k|}{|\calG|} 
\sum_{i,j\in\calG_k:\;  j\neq i} 
\frac{1}{|\calG_k|(|\calG_k|-1)} \Pr(\calE_{ij}|\bx_i,\bz)\nonumber\\
&\geq A_n\cdot 
\sum_{k=1}^N\frac{|\calG_k|}{|\calG|} 
\sum_{i,j\in\calG_k:\;  j\neq i} 
\frac{1}{|\calG_k|(|\calG_k|-1)} \cdot e^{-n\Omega^q_n(\widehat{P}_{\bx_i\bz\bx_j}, {W}^{{\delta}}_{Y|XZ})}\\
&\geq \frac{1}{2}A_n\cdot 
\sum_{k=1}^N\frac{|\calG_k|}{|\calG|} 
\sum_{i,j\in\calG_k:\;  j\neq i} 
\frac{1}{|\calG_k|(|\calG_k|-1)} \cdot e^{-n\min\{\Omega^q_n(\widehat{P}_{\bx_i\bz\bx_j}, {W}^{{\delta}}_{Y|XZ}),\Omega^q_n(\widehat{P}_{\bx_j\bz\bx_i}, {W}^{{\delta}}_{Y|XZ})\}}\label{eq: adgifdviuhf}\\
&\geq \frac{1}{2}A_n\cdot
\sum_{k=1}^N\frac{|\calG_k|}{|\calG|} 
\cdot e^{-n\left[\sum_{i,j\in\calG_k:\;  j\neq i} 
\frac{1}{|\calG_k|(|\calG_k|-1)} \min\left\{\Omega^q_n(\widehat{P}_{\bx_i\bz\bx_j}, {W}^{{\delta}}_{Y|XZ}),\Omega^q_n(\widehat{P}_{\bx_j\bz\bx_i}, {W}^{{\delta}}_{Y|XZ})\right\}\right]}\label{eq: aihv;oiduhfov;ihdfov}
\end{flalign}
where (\ref{eq: adgifdviuhf}) follows since $\frac{a+b}{2}\geq\frac{1}{2}\max\{a,b\}$, and 
(\ref{eq: aihv;oiduhfov;ihdfov}) follows from Jensen's Inequality applied to the convex function $f(t)=e^{-t}$. 

Next we treat the term which appears in the above square brackets of (\ref{eq: aihv;oiduhfov;ihdfov}). 
For convenience, denote by $\bx(t)$ the $t-th$ entry of the $n$-vector $\bx$. 
For any collection of codewords $\calL'\subseteq\calL$ such that $|\calL'|\geq 2$, 
we define the following distribution $\overline{P}^{\calL'}_{TZX\widetilde{X}}$. Let $T$ be a random variable uniformly distributed over $\{1,...,n\}$ and define
\begin{flalign}
\overline{P}^{\calL'}_{TZX\widetilde{X}}(t,z,x,\widetilde{x})&\triangleq 
\frac{1}{n}\cdot \indicator_{\{z=\bz(t)\}}\cdot \frac{\sum_{j\in \calL'}\indicator_{\{\bx_j(t)=x\}}}{ |\calL'|}\cdot 
\frac{\sum_{j\in \calL'}\indicator_{\{\bx_j(t)=\widetilde{x}\}}}{ |\calL'|}
.\label{eq: iusdgicugd}
\end{flalign}

Consider the function $\Omega^q(P_{XZ\widetilde{X}}, {W}^{{\delta}}_{Y|XZ})$ 
which extends $\Omega^q_n(P_{XZ\widetilde{X}}, {W}^{{\delta}}_{Y|XZ})$ (see definition in (\ref{eq: Ialpha dfndfugilugi})-(\ref{eq: calSdfn})) in a twofold manner: (a) it is defined for $P_{XZ\widetilde{X}}\in \calP(\calX^2\times \calZ)$ which need not necessarily be an empirical distribution of order $n$, and (b) the minimization is over the simplex $\calP(\calX^2\times\calZ\times\calY)$ rather than empirical distributions; that is, 
\begin{flalign}
\Omega^q(P_{XZ\widetilde{X}}, {W}^{{\delta}}_{Y|XZ})
&\triangleq 
\min_{V_{\widetilde{X}XYZ}\in\calS_q(P_{XZ\widetilde{X}})
 } D(V_{Y|XZ\widetilde{X}}\|{W}^{{\delta}}_{Y|XZ}|P_{XZ\widetilde{X}}) 
 .\label{eq: Ialpha dfndfugilugi11}
\end{flalign}

%
%

The following lemma is based on Plotkin's counting trick and the aforementioned Blinkovsky's idea similar to \cite{SHANNONGallagerBerlekamp1967522,BondaschiGuilleniFabregasDalai-IT2021}. 
\begin{lemma}\label{lm: audvgliudgvliu}
If $(W_{Y|X},q)$ is balanced, then for any $W_{Z|XY}$ 
any $\bz\in\calT_n(\hat{P}_Z)$, and any collection of codewords $\calB$ which is $(t,\widehat{P}_{XZ})$-quasi-symmetric w.r.t.\ $\bz$, it holds that:
\begin{flalign}
&\frac{1}{|\calB|(|\calB|-1)}\sum_{i,j\in \calB ,\; j\neq i} \min\left\{\Omega^q_n(\widehat{P}_{\bx_i\bz\bx_j}, {W}^{{\delta}}_{Y|XZ}),\Omega^q_n(\widehat{P}_{\bx_j\bz\bx_i}, {W}^{{\delta}}_{Y|XZ})\right\}\nonumber\\
&\leq 
\frac{|\calB|}{|\calB|-1}
\Omega^q\left(\overline{P}^{\calB}_{XZ\widetilde{X}},{W}^{{\delta}}_{Y|XZ}\right)+\overline{\zeta}_{t,{\delta},|\calB|}
\label{eq: afojb}
\end{flalign}
where 
${W}^{{\delta}}_{Y|XZ}$ is the marginal of $W_{Y|X}\times {W}^{{\delta}}_{Z|XY}$, with ${W}^{{\delta}}_{Z|XY}$ denoting the function of $(W_{Z|XY},{\delta})$ defined in (\ref{eq: ahfuv W star}),  
and $ \overline{\zeta}_{t,{\delta},|\calB|}=5\widetilde{K}({\delta})\cdot \left(
6/\sqrt{|\calB|}+2\sqrt{2/t}+3/t\right)$, where $\widetilde{K}({\delta})$ is a function of ${\delta}$. 
 \end{lemma}
Lemma \ref{lm: audvgliudgvliu} is proved in Appendix \ref{sc: fa;ihvoidfhiovhdfv}.

From Lemma \ref{lm: audvgliudgvliu} applied to $\calB=\calG_k$, (\ref{eq: alfgvilufdgv}), and (\ref{eq: aihv;oiduhfov;ihdfov}), it follows that 
\begin{flalign}
\Pr(error|\bz,\widehat{P}_{\bX\bZ}=\widehat{P}_{XZ})\geq&\frac{1}{|\calL|}\sum_{i\in\calL} \max_{j\in\calL:\; j\neq i}
\Pr(\calE_{ij}|\bx_i,\bz)\nonumber\\
&\geq \frac{1}{4}A_n\cdot
\sum_{k=1}^N\frac{|\calG_k|}{|\calG|} 
\cdot e^{-n\left[\frac{|\calG_k|}{|\calG_k|-1}\cdot \Omega^q\left(\overline{P}^{\calG_k}_{XZ\widetilde{X}}, {W}^{{\delta}}_{Y|XZ}\right)+\overline{\zeta}_{t,{\delta},|\calG_k|}\right]}\\
&\geq \frac{1}{4}A_n\cdot 
 e^{-n\max_k\left[\frac{M'}{M'-1}\cdot \Omega^q\left(\overline{P}^{\calG_k}_{XZ\widetilde{X}}, {W}^{{\delta}}_{Y|XZ}\right)+\overline{\zeta}_{t,{\delta},M'}\right]}
,\label{eq: v;ihdfov}
\end{flalign}
where (\ref{eq: v;ihdfov}) follows since by our construction $\forall k$, $|\calG_k|=M'$. 

The following lemma is taken from \cite[Lemma 13]{SomekhBaruchArxiv_16March2022}, which shows that the random variable $T$ in (\ref{eq: iusdgicugd}) can be replaced by another random variable $U$ of finite alphabet cardinality that does not increase with $n$ (as opposed to $T$). For the sake of completeness we include the proof of Lemma \ref{lm: Caratheodory} in Appendix \ref{ap: Caratheodory}. 
\begin{lemma}\label{lm: Caratheodory}
There exists a random variable $U$ whose alphabet size is $|\calU|\leq |\calX|^2|\calZ|$, and a joint distribution $P_{UXZ}\in\calP(\calU\times \calX\times\calZ)$ such that for any $(x,\widetilde{x},z)$, 
\begin{flalign}\label{eq: iuafgivufigvdfiu}
\sum_u P_{UZ}(u,z)P_{X|UZ}(x|u,z)P_{X|UZ}(\widetilde{x}|u,z)&=\overline{P}^{\calG_k}_{XZ\widetilde{X}}(x,z,\widetilde{x}).
\end{flalign}
There also exists a joint distribution $P_{UXZ}\in\calP(\calU\times \calX\times\calZ)$ such that for any $(x,z,\widetilde{x})$ (\ref{eq: iuafgivufigvdfiu}) holds, $Z$ is a deterministic function of $U$, and $|\calU|\leq |\calX|^2|\calZ|+1$.
\end{lemma}
Consider the set of distributions $P_{X\widetilde{X}Z}$ that can be expressed in the form 
\begin{flalign}
P_{X\widetilde{X}Z}(x,\widetilde{x},z)= \sum_u P_{XZ}(x,z)P_{U|XZ}(u|x,z)P_{X|ZU}(\widetilde{x}|z,u),\label{eq: aiugvliufv}
\end{flalign}
 and whose marginals are both equal to $\widehat{P}_{XZ}$ ; that is,
\begin{flalign}
\calQ(\widehat{P}_{XZ})\triangleq \left\{V_{XZ\widetilde{X}}:\;\substack{V_{XZ}=V_{\widetilde{X}Z}=\widehat{P}_{XZ},\\ \exists V_{U|XZ}:\forall (x,\widetilde{x},z),\;V_{XZ\widetilde{X}}(x,z,\widetilde{x})= \sum_u V_{XZ}(x,z)V_{U|XZ}(u|x,z)V_{X|ZU}(\widetilde{x}|z,u)}
\right\}.
\end{flalign}
Hence, 
\begin{flalign}
&\max_k \Omega^q\left(\overline{P}^{\calG_k}_{XZ\widetilde{X}},  {W}^{{\delta}}_{Y|XZ}\right)
\leq \sup_{P_{XZ\widetilde{X}}\in \calQ(\widehat{P}_{XZ})}\Omega^q\left(P_{XZ\widetilde{X}},  {W}^{{\delta}}_{Y|XZ}\right).\label{eq: aiufgviufdgv}
\end{flalign}

We next claim that the supremum in (\ref{eq: aiufgviufdgv}) can be replaced with a maximum. 
 This follows since (a) Lemma \ref{lm: Caratheodory} implies that 
the set $\calQ(\widehat{P}_{XZ})$ is closed and convex (and bounded in $\mathbb{R}^{|\calX|^2|\calZ|}$),  
(b) Lemma \ref{lm: continuity sym result} to follow shows that the function $\Omega^q(P_{XZ\widetilde{X}}, W_{Y|XZ}^{\delta})$ is continuous for symmetric distributions; i.e., for $P_{XZ\widetilde{X}}\in \calP_{sym}(\calX^2\times \calZ)$, where $\calP_{sym}(\calX^2\times \calZ)$ is defined in (\ref{eq: sym dfn set}),  
and (c) the supremum of a continuous function over a convex compact set is attained.

\begin{lemma}\label{lm: continuity sym result}
If the pair $(q,\overline{W}_{Y|XZ})$ is balanced, then $\Omega^q_q(P_{XZ\widetilde{X}},\overline{W}_{Y|XZ})$ is a continuous function of $P_{XZ\widetilde{X}}$ over the set $\calP_{sym}(\calX^2\times \calZ)$. 
\end{lemma}
Lemma \ref{lm: continuity sym result} is proved in Appendix \ref{sc: Proof of continuity sym lemma}. 
Now, with a slight abuse of notation, we denote\footnote{In fact, $
\widetilde{\Omega^q}(P_{XZU}, {W}^{{\delta}}_{Y|XZ})= \eta_q(P_{XZU},W_{Y|XZ})$, where $\eta_q(P_{XZU},W_{Y|XZ})$ is defined in (\ref{eq: aiugvliufdglivugdfuiv}), due to the dual form representation (\ref{eq: afovfihosbjpfogjbposjgfbojgojfpjb}).}
\begin{flalign}\label{eq: Omega tag dfn}
\widetilde{\Omega^q}(P_{XZU}, {W}^{{\delta}}_{Y|XZ})&= \Omega^q(P_{XZ\widetilde{X}},  {W}^{{\delta}}_{Y|XZ}),
\end{flalign}
where $P_{X\widetilde{X}Z}$ and $P_{U|XZ}$ are related through (\ref{eq: aiugvliufv}).

Thus, applying Lemma \ref{lm: continuity sym result} to $\overline{W}= W^{\delta}$, we have established that 
\begin{flalign}
&\max_k \Omega^q\left(\overline{P}^{\calG_k}_{XZ\widetilde{X}},  {W}^{{\delta}}_{Y|XZ}\right)\nonumber\\
&\leq \max_{P_{XZ\widetilde{X}}\in \calQ(\widehat{P}_{XZ})}\Omega^q\left(P_{XZ\widetilde{X}},  {W}^{{\delta}}_{Y|XZ}\right)\label{eq: aiufghvidufhivudhf}\\
&\triangleq \max_{P_{U|XZ}}\widetilde{\Omega^q}(\widehat{P}_{XZ}\times P_{U|XZ},  {W}^{{\delta}}_{Y|XZ})\label{eq: adfhv;iudfhv;iodhf;}.
\end{flalign}
Consequently, whenever $|\calL(\bz,\widehat{P}_{XZ})|\geq e^{n\hat{\epsilon}_n}$,
\begin{flalign}
\Pr(error|\bz,\widehat{P}_{\bX\bZ}=\widehat{P}_{XZ})&\geq 
 \frac{A_n}{4}\cdot 
 e^{-n\max_k\left[\frac{M'}{M'-1}\cdot \Omega^q\left(\overline{P}^{\calG_k}_{XZ\widetilde{X}},  {W}^{{\delta}}_{Y|XZ}\right)+\overline{\zeta}_{t,{\delta},M'}\right]}\\
 &\geq 
 \frac{A_n}{4}
\cdot e^{-n\left[\frac{M'}{M'-1}\cdot\max_{P_{U|XZ}}\widetilde{\Omega^q}(\hat{P}_{XZ}\times P_{U|XZ},  {W}^{{\delta}}_{Y|XZ})
+\overline{\zeta}_{t,{\delta},M'}\right]}.\label{eq: afiud;hvv}
\end{flalign}

The following lemma was proved in \cite{SomekhBaruchArxiv_16March2022} (see Lemma 4 therein). For the sake of completeness, the proof is included in Appendix \ref{sc: LIST  size lemma Proof}. 
\begin{lemma}\label{eq: List size lemma}
Let a codebook $\calC_n=\{\overline{\bx}_i\}_{i=1}^{\mathbb{M}_n}$ be given, let $\bX$ denote the random codeword (distributed uniformly over $\calC_n$), and let $\bZ$ denote the output of the channel $ {W}^{{\delta}}_{Z|X}$ when fed by $\bX$. 
For any $\tau\geq 0$, 
\begin{flalign}\label{eq: sdhvodishfiov}
&\Pr\left(
 |\calL(\bZ, \widehat{P}_{XZ})|\geq e^{n\tau}\big|\widehat{P}_{\bX\bZ}=\widehat{P}_{XZ}\right)\geq 1-(n+1)^{|\calX||\calZ|-1}\cdot e^{-n[R-I(\widehat{P}_{XZ})-\tau]}.
\end{flalign}
\end{lemma}

From Lemma \ref{eq: List size lemma}, we deduce that 
for any $\widehat{P}_{XZ}$, $\widetilde{\epsilon}_n>0$, and $\hat{\epsilon}_n>0$, such that 
\begin{flalign}\label{eq: ilsufgviludf}
R\geq I(\widehat{P}_{XZ})+\hat{\epsilon}_n+\frac{|\calX||\calZ|-1}{n}\log(n+1)+\widetilde{\epsilon}_n,
\end{flalign}
it holds that
$\Pr\left(|\calL(\bZ,\widehat{P}_{XZ})|<e^{n\hat{\epsilon}_n}\big|\widehat{P}_{\bX\bZ}=\widehat{P}_{XZ}\right)\leq  e^{-n\widetilde{\epsilon}_n} $.
Consequently, 
for $\widetilde{\epsilon}_n>1/n$, we have
\begin{flalign}
\Pr(error|\widehat{P}_{XZ})&\geq \Pr(error, \calL(\bZ,\widehat{P}_{XZ})\geq e^{n\hat{\epsilon}_n}|\widehat{P}_{\bX\bZ}=\widehat{P}_{XZ})\label{eq: aiufviufgv}\\
&\geq \left(1-e^{-n\widetilde{\epsilon}_n}\right)\cdot 
\Pr(error|\widehat{P}_{\bX\bZ}=\widehat{P}_{XZ},\calL(\bZ,\widehat{P}_{XZ})\geq e^{n\hat{\epsilon}_n})\\
&\geq \left(1-e^{-n\widetilde{\epsilon}_n}\right)\cdot 
\min_{\bz\in\calT_n(\widehat{P}_Z):\; |\calL|
\geq e^{n\hat{\epsilon}_n}}\Pr(error|\widehat{P}_{\bX\bZ}=\widehat{P}_{XZ},\bZ=\bz)\label{eq: udfhv0}
\\
&\geq \left(1-e^{-n\widetilde{\epsilon}_n}\right)\cdot 
\frac{A_n}{4}
\cdot e^{-n\left[\frac{M'}{M'-1}\cdot\max_{P_{U|XZ}}\widetilde{\Omega^q}(\hat{P}_{XZ}\times P_{U|XZ},  {W}^{{\delta}}_{Y|XZ})
+\overline{\zeta}_{t,{\delta},M'}\right]}
,\label{eq: udfhv}
\end{flalign}
where the last step follows from (\ref{eq: afiud;hvv}). 

For $M'$ arbitrarily large, recall we take $n$ such that (\ref{eq: avuh;difuhv;oihdf;oiv}) holds, 
ensuring that the conditions for Lemma \ref{lm: Komlos Lemma generalized second lemma} hold. 
Note that (\ref{eq: udfhv}) holds for every $ \widehat{P}_{XZ}\in\calP_n(\calX\times \calZ)$ such that $\widehat{P}_{X}=P_n$, and which is a possible empirical distribution of $\bX,\bZ$, and which satisfies (\ref{eq: ilsufgviludf}). Therefore, 
\begin{flalign}
\Pr(error)&=\sum_{\hat{P}_{XZ}:\; \hat{P}_{X}=P_n}  \Pr
(\hat{P}_{\bX\bZ}=\hat{P}_{XZ})\cdot \Pr(error|\widehat{P}_{XZ}) \\
&\geq\sum_{\widehat{P}_{XZ}:\; \widehat{P}_X=P_n} \frac{1}{(n+1)^{|\calX||\calZ|}}e^{-nD(\widehat{P}_{Z|X}\| {W}^{{\delta}}_{Z|X}|\widehat{P}_X)}\cdot \Pr(error|\widehat{P}_{XZ}) ,\label{eq: aiuvgiufgv}
\end{flalign} 
and denoting $\epsilon_{n,t,{\delta},M'}=\frac{1}{n}[\log 4-\log (1-e^{-n\widetilde{\epsilon}_n}) ]+\frac{|\calX||\calZ|}{n}\log(n+1)+\overline{\zeta}_{t,{\delta},M'}$, we get
\begin{flalign}\label{eq: idfugvsliudfg}
&-\frac{1}{n}\log \Pr(error)\leq \nonumber\\
&\min_{\substack{
\widehat{P}_{XZ}\in\calP_n(\calX\times \calZ):\; \widehat{P}_{X}=P_n,\\
I(\widehat{P}_{XZ})\leq R-\epsilon_n''}}D(\widehat{P}_{Z|X}\| {W}^{{\delta}}_{Z|X}|\widehat{P}_X)+\frac{M'}{M'-1}\max_{P_{U|XZ}}\widetilde{\Omega^q}(\hat{P}_{XZ}\times P_{U|XZ},  {W}^{{\delta}}_{Y|XZ})
+\epsilon_{n,t,{\delta},M'}.\\
&\leq \frac{M'}{M'-1}\bigg[\min_{\substack{
\widehat{P}_{XZ}\in\calP_n(\calX\times \calZ):\; \widehat{P}_{X}=P_n,\\
I(\widehat{P}_{XZ})\leq R-\epsilon_n''}}D(\widehat{P}_{Z|X}\| {W}^{{\delta}}_{Z|X}|\widehat{P}_X)+\max_{P_{U|XZ}}\widetilde{\Omega^q}(\hat{P}_{XZ}\times P_{U|XZ},  {W}^{{\delta}}_{Y|XZ})
+\epsilon_{n,t,{\delta},M'}.\bigg],
\end{flalign}
where $\epsilon_n''\triangleq \hat{\epsilon}_n+\frac{|\calX||\calZ|-1}{n}\log(n+1)+\widetilde{\epsilon}_n$.
The following lemma shows that the minimization over empirical distributions $\calP_n(\calX\times \calZ)$ can be approximated by a minimization over the simplex $\calP(\calX\times \calZ)$. 
\begin{lemma}\label{lm: technical ahuv;oiuhdfiouh}
For any ${\delta}>0$, there exist vanishing sequences $\epsilon_{a,n}$, $\epsilon_{b,n}$ such that for all $R>0$ and any ${W}^{{\delta}}_{Z|XY}$ which satisfies 
${W}^{{\delta}}_{Z|XY}(z|x,y)\geq \frac{{\delta}}{|\calZ|}$ 
\begin{flalign}
&\min_{\substack{
\widehat{P}_{XZ}\in\calP_n(\calX\times \calZ):\; \widehat{P}_{X}=P_n,\\
I(\widehat{P}_{XZ})\leq R+\epsilon_{a,n}}}D(\widehat{P}_{Z|X}\| {W}^{{\delta}}_{Z|X}|\widehat{P}_X)+\max_{P_{U|XZ}}\widetilde{\Omega^q}(\hat{P}_{XZ}\times P_{U|XZ},  {W}^{{\delta}}_{Y|XZ})-\epsilon_{b,n}\nonumber\\
&\leq 
\min_{\substack{
P_{XZ}\in\calP(\calX\times \calZ):\; P_{X}=P_n,\\
I(P_{XZ})\leq R}}D(P_{Z|X}\| {W}^{{\delta}}_{Z|X}|P_X)+\max_{P_{U|XZ}}\widetilde{\Omega^q}(P_{XZU},  {W}^{{\delta}}_{Y|XZ}).\label{eq: auvghiuhdfgv}
\end{flalign}
\end{lemma}
The proof of the Lemma \ref{lm: technical ahuv;oiuhdfiouh} appears in Appendix \ref{sc: Proof tec}. 
Further, by definition of $\widetilde{\Omega^q}(P_{XZU},  {W}^{{\delta}}_{Y|XZ})$, we have
\begin{flalign}
& \min_{\substack{
P_{XZ}\in\calP(\calX\times \calZ):\\ P_{X}=P_n,\\
I(P_{XZ})\leq R}}\max_{P_{U|XZ}} D(P_{Z|X}\| {W}^{{\delta}}_{Z|X}|P_X)+\widetilde{\Omega^q}(P_{XZU}, {W}^{{\delta}}_{Y|XZ})\nonumber\\
& =\min_{\substack{
P_{XZ}\in\calP(\calX\times \calZ):\\ P_{X}=P_n,\\
I(P_{XZ})\leq R}}\max_{P_{U|XZ}} 
\min_{P_{Y|XZ\widetilde{X}} \in\calS_q^{cond}(P_{XZ\widetilde{X}}) }
D(P_{YZ|X}\|{W}^{{\delta}}_{YZ|X}|P_X)+I(\widetilde{X};Y|X,Z),\label{eq: aiufgviufdgviud}
\end{flalign}
where $P_{X\widetilde{X}Z}(x,\widetilde{x},z)=\sum_u P_{XZU}(x,z,u)P_{X|UZ}(\widetilde{x}|u,z)$.

We next wish to replace ${W}^{{\delta}}_{YZ|X}$ in the above divergence by $W_{YZ|X}$ to establish the desired bound. 
Recalling that ${W}^{{\delta}}_{Z|XY}=(1-{\delta})W_{Z|XY}+{\delta} \widetilde{Q}_Z$, where $\widetilde{Q}_Z(z)=\frac{1}{|\calZ|}$, we have 
${W}^{{\delta}}_{YZ|X}=(1-{\delta})W_{Y|X}\times W_{Z|XY}+{\delta} W_{Y|X}\times \widetilde{Q}_Z$, which yields
\begin{flalign}
D(P_{YZ|X}\|{W}^{{\delta}}_{YZ|X}|P_X)&\leq (1-{\delta})D(P_{YZ|X}\|W_{Y|X}\times W_{Z|XY}|P_X)+
{\delta}D(P_{YZ|X}\|W_{Y|X}\times \widetilde{Q}_Z|P_X).
\end{flalign}
Since $D(P_{YZ|X}\|W_{Y|X}\times \widetilde{Q}_Z|P_X)=D(P_{Y|X}\|W_{Y|X}|P_X)+D(P_{Z|XY}\|\widetilde{Q}_Z|P_{XY})$, and $D(P_{Z|XY}\|\widetilde{Q}_Z|P_{XY})\leq \log|\calZ|$, this gives
\begin{flalign}
& \min_{\substack{
P_{XZ}\in\calP(\calX\times \calZ):\\ P_{X}=P_n,\\
I(P_{XZ})\leq R}}\max_{P_{U|XZ}} D(P_{Z|X}\|{W}^{{\delta}}_{Z|X}|P_X)+\widetilde{\Omega^q}(P_{XZU}, {W}^{{\delta}}_{Y|XZ})\nonumber\\
& \leq \min_{\substack{
P_{XZ}\in\calP(\calX\times \calZ):\\ P_{X}=P_n,\\
I(P_{XZ})\leq R}}\max_{P_{U|XZ}} 
\min_{P_{Y|XZ\widetilde{X}} \in\calS_q^{cond}(P_{XZ\widetilde{X}}) }
\bigg[D(P_{YZ|X}\| W_{YZ|X}|P_X)+I(\widetilde{X};Y|X,Z)\nonumber\\
&+ {\delta}\left(
D(P_{Y|X}\|W_{Y|X}|P_X)+\log|\calZ|\right)\bigg]\\
& \leq (1+{\delta}) \min_{\substack{
P_{XZ}\in\calP(\calX\times \calZ):\\ P_{X}=P_n,\\
I(P_{XZ})\leq R}}\max_{P_{U|XZ}} 
\min_{P_{Y|XZ\widetilde{X}} \in\calS_q^{cond}(P_{XZ\widetilde{X}}) }
\bigg[D(P_{YZ|X}\|W_{YZ|X}|P_X)+I(\widetilde{X};Y|X,Z)\bigg]\nonumber\\
&+ {\delta}
\log|\calZ|,
\end{flalign}
where the last step follows since $D(P_{YZ|X}\|W_{YZ|X}|P_X)\geq D(P_{Y|X}\|W_{Y|X}|P_X)$. 

To conclude, if $(q,W_{Y|X})$ is balanced, and $n$ is such that $ e^{n\hat{\epsilon}_n}>4\cdot \max\left\{M_0(M',t),M'\right\}$ (see (\ref{eq: avuh;difuhv;oihdf;oiv})), 
then for any $W_{Z|XY}$ we have
\begin{flalign}
P(error)&\geq e^{-n(1+{\delta}_n) \frac{M'}{M'-1}\left[\overline{E}^q(R-\epsilon_n''-\epsilon_{a,n},P_n,W_{YZ|X})+
\epsilon_{n,t,{\delta},M'}.+\epsilon_{b,n}+ {\delta}\log|\calZ|\right]}.
\end{flalign}
Since $\overline{E}^q(R,P_n,W_{YZ|X})$ is bounded (e.g., by $\overline{E}^q(R,P_n,W_{YZ|X})|_{R=0}$, which can be assumed bounded, as otherwise (\ref{eq: dfhvdiudsvhidfiludffdhliufihi}) follows straightforwardly), and since ${\delta}$ can be taken to be arbitrarily small, and accordingly, $M',t,n$ can be taken to be arbitrarily large, there exists a vanishing sequence $\epsilon_n$, such that 
$$P(error)\geq e^{-n\left(\overline{E}^q(R-\epsilon_n,P_n,W_{YZ|X})+\epsilon\right)}.$$

\section{Concluding Remarks}\label{sc: Conclusion}

In this paper we presented a new bound on the reliability function of the DMC. 
Unlike previous analyses, where separate derivations were performed for zero rate and for higher rates, and then an argument to establish the linear straight line bound between the two was used for the ML decoding metric, 
our main result is obtained in a unified manner, which is valid for any rate and a wide class of decoding metrics including ML. 

We showed that our bound is at least as tight as former bounds: the sphere-packing bound, the zero-rate reliability function, the straight line bound, and an amendment of Blahut's bound. We also established that in certain cases, there is a {\it strict inequality} between our new bound and these former results for a certain range of low rates. 

The proof technique presented in this paper relied on a genie aided receiver of a specific kind, that is provided with a narrowed list of codewords to choose from. This technique already turned out to be very useful in deriving the best known single-letter upper bound on the mismatch capacity and bounds on the mismatched reliability function, but it is more surprising to see that it actually yields strictly tighter results compared to previous works also in the maximum-likelihood decoding case. 

\appendix

\subsection{Proof of Proposition \ref{pr: dual}}\label{ap: DUAL EXP APPENDIX}


Recall that we consider additive $q\in\mathbb{R}\cup\{-\infty\}$. 
The proposition follows since for any fixed $P_{XZ\widetilde{X}}$, it holds that 
\begin{flalign}
&\Omega^q(P_{XZ\widetilde{X}}, W_{Y|XZ})\nonumber\\
&=\sup_{s\geq 0}-\sum_{x,z\widetilde{x}}P_{XZ\widetilde{X}}(x,z,\widetilde{x})\log \left(\sum_yW_{Y|XZ}(y|x,z)e^{s[q(\widetilde{x},y)-q(x,y)]}\right),\label{eq: afovfihosbjpfogjbposjgfbojgojfpjb}
\end{flalign}
where (\ref{eq: afovfihosbjpfogjbposjgfbojgojfpjb}) follows verbatim as the line of proof of \cite[Eqs.\ (106)-(111)]{ScarlettPengMerhavMartinezGuilleniFabregas_mismatch_2014_IT} with $P_{X\widetilde{X}Z}$ in the role of $P_{X\widetilde{X}}$, $(x,z)$ and $(\widetilde{x},z)$ in the roles of $x$ and $\widetilde{x}$,  respectively, and for any $x,y,z$ we define $q((x,z),y)\triangleq q(x,y)$.
Further, an additive metric $q\in\mathbb{R}\cup\{-\infty\}$ is equivalent to a multiplicative non-negative metric (considered in \cite{ScarlettPengMerhavMartinezGuilleniFabregas_mismatch_2014_IT}) by taking the logarithm. 
Finally, in the additive metric case $e^{s[q(\widetilde{x},y)-q(x,y)]}$ translates to $\left(\frac{q(\widetilde{x},y)}{q(x,y)}\right)^s$ in the multiplicative case.

\subsection{Proof of Theorem \ref{th: strict inprovement thm}}\label{sc: strict inprovement thm}

For any $\alpha\in[0,1]$ we use the shorthand notation $\overline{\alpha}=1-\alpha$. 
To prove (\ref{eq: sl proof}), note that (\ref{eq: W chann alpha dfn}) implies that $W_{Z|X}^{(\alpha)}(z|x)= \alpha\cdot W_{Y|X}(z|x)+\overline{\alpha}\cdot Q_Z(z)$, and 
$W_{Y|XZ}^{(\alpha)}(y|x,z)=\alpha\cdot\indicator_{\{y=z\}}+\overline{\alpha}\cdot W_{Y|X}(y|x)$, and thus from (\ref{eq: afouhvuidfhvadv}) we get
\begin{flalign}
&\overline{E}^{q}(P_{XZ},W_{YZ|X}^{(\alpha)}) \nonumber\\
& = D(P_{Z|X}\|W_{Z|X}^{(\alpha)}|P)+\max_{P_{U|XZ}}
\min_{ \substack{ P_{Y|XZ\widetilde{X}}:\\ 
\EE_P q(\widetilde{X},Y)\geq \EE_P q(X,Y)
} }\phi(P_{XZU},P_{Y|XZ\widetilde{X}},W_{Y|XZ}^{(\alpha)}),\end{flalign}
where the expectations $\EE_P q(\widetilde{X},Y)$, and $\EE_P q(X,Y)$ are w.r.t.\ 

\noindent $P_{XZU\widetilde{X}Y}(x,z,u,\widetilde{x},y)=P_{ZXU}(x,z,u)P_{X|ZU}(\widetilde{x}|z,u)P_{Y|XZ\widetilde{X}}(y|x,z,\widetilde{x})
$.

The rightmost term inside the maximization over $P_{U|XZ}$ satisfies:
\begin{flalign}
&\min_{ \substack{ P_{Y|XZ\widetilde{X}}:\\ 
\EE_P q(\widetilde{X},Y)\geq \EE_P q(X,Y)
} }\phi(P_{XZU},P_{Y|XZ\widetilde{X}},W_{Y|XZ}^{(\alpha)})\nonumber\\
& \leq 
\min_{ \substack{ V_{Y|XZ\widetilde{X}}:\\ 
\EE_V q(\widetilde{X},Y)\geq \EE_V q(X,Y)
} }\phi(P_{XZU}, \alpha\cdot\indicator_{\{Y=Z\}}+\overline{\alpha}\cdot  V_{Y|XZ\widetilde{X}},W_{Y|XZ}^{(\alpha)})\label{eq: aiufgviudgfiv}\\
& \leq 
\min_{ \substack{V_{Y|XZ\widetilde{X}}:\\ 
\EE_V q(\widetilde{X},Y)\geq \EE_V q(X,Y)
} }\alpha\cdot \phi(P_{XZU},\indicator_{\{Y=Z\}}, \indicator_{\{Y=Z\}})
+\overline{\alpha} \cdot \phi(P_{XZU},V_{Y|XZ\widetilde{X}}, W_{Y|X})\label{eq: aofuhv;ufhv}\\
& =  \overline{\alpha} \cdot
\min_{ \substack{V_{Y|XZ\widetilde{X}}:\\ 
\EE_V q(\widetilde{X},Y)\geq \EE_V q(X,Y)
} }\phi(P_{XZU},V_{Y|XZ\widetilde{X}}, W_{Y|X})\label{eq: adoivhodfihvohfdo}\\
& \leq   \overline{\alpha}\cdot  
\min_{ \substack{V_{Y|XZ\widetilde{X}}:\\ Z-(X,\widetilde{X})-Y\\
\EE_V q(\widetilde{X},Y)\geq \EE_V q(X,Y)
} } \phi(P_{XZU},V_{Y|XZ\widetilde{X}}, W_{Y|X}),
\end{flalign}
where the expectations $\EE_V q(\widetilde{X},Y)$, and $\EE_V q(X,Y)$ are w.r.t.\ 

\noindent $V_{XZU\widetilde{X}Y}(x,z,u\widetilde{x},y)=P_{ZXU}(x,z,u)P_{X|ZU}(\widetilde{x}|z,u)V_{Y|XZ\widetilde{X}}(y|x,z,\widetilde{x})
$, (\ref{eq: aiufgviudgfiv}) follows by taking $P_{Y|XZ\widetilde{X}}$ of the form $ \alpha\cdot\indicator_{\{Y=Z\}}+\overline{\alpha}\cdot  V_{Y|XZ\widetilde{X}}$ and since $P_{XZ}=P_{\widetilde{X}Z}$, one has 
$\EE q(\widetilde{X},Z)\geq \EE q(X,Z)$, so by linearity of the expectation, we have $\EE_P q(\widetilde{X},Y)- \EE_P q(X,Y)=\EE_V q(\widetilde{X},Y)- \EE_V q(X,Y)$. 
The inequality (\ref{eq: aofuhv;ufhv}) follows from convexity of $\phi(P_{XZU},P_{Y|XZ\widetilde{X}}, W_{Y|X})$ in $P_{Y|XZ\widetilde{X}}$ (see (\ref{eq: afouhvuidfhvadv})), further (\ref{eq: adoivhodfihvohfdo}) follows since $\phi(P_{XZU},\indicator_{\{Y=Z\}}, \indicator_{\{Y=Z\}})=0$, and 
the last inequality follows by adding the constraint $Z-(X,\widetilde{X})-Y$ to the minimization. 
Maximizing over $P_{U|XZ}$, we get
\begin{flalign}
& \max_{P_{U|XZ}}
\min_{ \substack{V_{Y|XZ\widetilde{X}}:\\ Z-(X,\widetilde{X})-Y\\
\EE_V q(\widetilde{X},Y)\geq \EE_V q(X,Y)
} } \phi(P_{XZU},V_{Y|XZ\widetilde{X}}, W_{Y|X})\\
&= \max_{\substack{V_{\widetilde{X}UXZ}:\\
V_{XZ}=P_{XZ}\\
V_{XZU}= V_{\widetilde{X}ZU}\\
\widetilde{X}-(U,Z)-X
}}
\min_{ \substack{\widetilde{V}_{X\widetilde{X}Y}:\\ \widetilde{V}_{X\widetilde{X}}=V_{X\widetilde{X}}\\
\EE_{\widetilde{V}} q(\widetilde{X},Y)\geq \EE_{\widetilde{V}} q(X,Y)
} }D(\widetilde{V}_{Y|X\widetilde{X}}\|W_{Y|X}|V_{X\widetilde{X}})\label{eq: a'ofivofdihvoijf}\\
&\leq  \max_{\substack{V_{\widetilde{X}U'|X}:\\
V_{X}=P_{X}\\
V_{XU'}= V_{\widetilde{X}U'}\\
\widetilde{X}-U'-X
}}
\min_{ \substack{\widetilde{V}_{X\widetilde{X}Y}:\\ \widetilde{V}_{X\widetilde{X}}=V_{X\widetilde{X}}\\
\EE_{\widetilde{V}} q(\widetilde{X},Y)\geq \EE_{\widetilde{V}} q(X,Y)
} }D(\widetilde{V}_{Y|X\widetilde{X}}\|W_{Y|X}|V_{X\widetilde{X}}),
\end{flalign}
where (\ref{eq: a'ofivofdihvoijf}) holds by definition of $\phi(P_{XZU},V_{Y|XZ\widetilde{X}}, W_{Y|X})$ for the case where $\widetilde{X}-(U,Z)-X$, and the last step follows by taking $U'=(U,Z)$, and by relaxing the constraint $V_{XZ}=P_{XZ}$ to the looser one $V_{X}=P_{X}$. Note that the last term on the r.h.s.\ equals $\overline{E}_0^q(P,W)$ (see (\ref{eq: aoufhv;ufdhv})), thus 
\begin{flalign}
&\min_{P_{Z|X}:\; I(P_{XZ})\leq \alpha R}\overline{E}^{q}(P_{XZ},W_{YZ|X}^{(\alpha)}) \nonumber\\
&\leq \min_{P_{Z|X}:\; I(P_{XZ})\leq \alpha R} D(P_{Z|X}\|W_{Z|X}^{(\alpha)}|P)+\overline{\alpha}\overline{E}_0^q(P,W).\label{eq: aoufhv;ufdhv2}
\end{flalign}
Since $\overline{E}_0^q(P,W)$ does not depend on $P_{Z|X}$, the minimization over $P_{Z|X}$ affects only the first term. 
The minimized first term satisfies 
\begin{flalign}
&\min_{P_{Z|X}:\; I(P\times P_{Z|X})\leq \alpha R} D(P_{Z|X}\|W_{Z|X}^{(\alpha)}|P)\nonumber\\
&\leq \min_{V_{Z|X}:\; I(P\times V_{Z|X} )\leq R} 
D(\overline{\alpha}\cdot Q_Z +\alpha \cdot V_{Z|X} \|W_{Z|X}^{(\alpha)}|P)\label{eq: ilufgviluzdf}\\
&\leq \min_{V_{Z|X}:\; I(P\times V_{Z|X} )\leq R}
\alpha \cdot D(V_{Z|X} \|W_{Y|X}|P)\label{eq: a;iufgviufdgv}
\end{flalign}
where (\ref{eq: ilufgviluzdf}) follows by convexity of $I(P_X\times P_{Z|X})$ in $P_{Z|X}$ for fixed $P_X$, which implies that $I(P_X\times[\overline{\alpha}\cdot Q_Z +\alpha \cdot V_{Z|X} ])\leq \overline{\alpha }I(P_X\times Q_Z)+ \alpha I(P_X\times V_{Z|X})=
\alpha I(P_X\times V_{Z|X})$, and (\ref{eq: a;iufgviufdgv}) follows due to the convexity of the divergence; i.e., $D(\alpha P_1+\overline{\alpha } P_2\|\alpha Q_1+\overline{\alpha } Q_2)\leq \alpha D(P_1\|P_2) + \overline{\alpha } D(P_2\|Q_2)$, and by definition of $W_{Z|X}^{(\alpha)}$ (\ref{eq: W chann alpha dfn}), which implies $W_{Z|X}^{(\alpha)}(z|x)= \alpha\cdot W_{Y|X}(z|x)+\overline{\alpha}\cdot Q_Z(z)$. 

Since $\calZ=\calY$, this concludes the proof of (\ref{eq: sl proof}). 
Taking $\alpha=1$ and maximizing over $P_X\in\calP(\calX)$ we obtain (\ref{eq: audfgviufdg}). 
Taking $\alpha=0$ and maximizing over $P_X\in\calP(\calX)$ we obtain 
\begin{flalign}
\max_{P\in\calP(\calX)} \overline{E}^{q}(R,P,W_{Y|X}\times Q_Z)\leq \max_{P\in\calP(\calX)}  \overline{E}_0^q(P,W)
\end{flalign}

Note that from the dual form (\ref{eq: afovfihosbjpfogjbposjgfbojgojfpjb}) (applied to null $Z$) we have
\begin{flalign}
&\max_{P}\overline{E}_0^q(P,W)\\
&=\max_{P_{XU}}\sup_{s\geq 0}
-\sum_{u,\widetilde{x},x}P_{XU}(x,u)P_{X|U}(\widetilde{x}|u)\log \sum_yW(y|x)e^{s[q(\widetilde{x},y)-q(x,y)]}\\
&\leq \max_{P_{XU}}\sup_{s\geq 0}\max_{u'\in\calU}
-\sum_{\widetilde{x},x}P_{X|U}(x|u')P_{X|U}(\widetilde{x}|u')\log \sum_yW(y|x)e^{s[q(\widetilde{x},y)-q(x,y)]}\\
&= \max_{P_X}\sup_{s\geq 0}
-\sum_{\widetilde{x},x}P_X(x)P_X(\widetilde{x})\log \sum_yW(y|x)e^{s[q(\widetilde{x},y)-q(x,y)]}\\
&= E_{ex}^q(0^+,W)
,
\end{flalign}
proving (\ref{eq: fiuvb}).

Now, we claim that for $R>0$, at least one of the two inequalities (\ref{eq: ilufgviluzdf})-(\ref{eq: a;iufgviufdgv}) is strict; that is, 
\begin{flalign}
&\min_{P_{Z|X}:\; I(P_{XZ})\leq \alpha R} D(P_{Z|X}\|W_{Z|X}^{(\alpha)}|P)\nonumber\\
&< \min_{V_{Z|X}:\; I(P_X\times V_{Z|X} )\leq R}
\alpha \cdot D(V_{Z|X} \|W_{Y|X}|P).\label{eq: a;iufgviufdgvavh;fd}
\end{flalign}

To prove this, first note that one can assume that $\min_{V_{Z|X}:\; I(P_X\times V_{Z|X} )\leq R}
\alpha \cdot D(V_{Z|X} \|W_{Y|X}|P)$ is finite, as otherwise the strict inequality follows trivially, since by definition of $W_{Z|X}^{(\alpha)}$, the l.h.s.\ expression $\min_{P_{Z|X}:\; I(P_{XZ})\leq \alpha R} D(P_{Z|X}\|W_{Z|X}^{(\alpha)}|P)$ is finite for $\alpha\in(0,1)$. 
Next, assume in negation that the two inequalities (\ref{eq: ilufgviluzdf})-(\ref{eq: a;iufgviufdgv}) are in fact equalities. Note that the function $\min_{V_{Z|X}:\; I(P\times V_{Z|X} )\leq R}
D(V_{Z|X} \|W_{Y|X}|P)$ is nothing but the sphere-packing bound for $P$-constant composition codes, which is well known to be a strictly decreasing function of $R$ (for the range where it is finite up to the point where it becomes zero, which equals $R=I(P\times W_{Y|X})$). Therefore, the constraint $I(P\times V_{Z|X} )\leq R$ in (\ref{eq: a;iufgviufdgv}) must be attained with equality. 
From the assumption (in negation) this must also hold for the expressions on both sides of (\ref{eq: ilufgviluzdf}); i.e., the constraints are met with equalities. 
This implies that we assume by contradiction that 
\begin{flalign}\label{eq: aiufhv;iufhd;ivhdfiohv;d}
I(P\times \widetilde{P}_{Z|X})<\alpha R\Rightarrow D(\widetilde{P}_{Z|X}\|W_{Z|X}^{(\alpha)}|P)>\min_{P_{Z|X}:\; I(P_{XZ})\leq \alpha R} D(P_{Z|X}\|W_{Z|X}^{(\alpha)}|P).
\end{flalign}

Letting $V_{Z|X}^*$ be the optimizer of the r.h.s.\ of (\ref{eq: a;iufgviufdgv}), it can be easily shown that unless $V_{Z|X}^*$ is such that $I(P_X\times V_{Z|X}^*)=0$, we have that $I(P_X\times[\overline{\alpha}\cdot Q_Z +\alpha \cdot V_{Z|X}^* ])$ is strictly convex, i.e.,
\begin{flalign}
I(P_X\times[\overline{\alpha}\cdot Q_Z +\alpha \cdot V_{Z|X}^* ])< \overline{\alpha }I(P_X\times Q_Z)+ \alpha I(P_X\times V_{Z|X}^*)=\alpha I(P_X\times V_{Z|X}^*).\label{eq: ads;oh;odivi}
\end{flalign}
To establish this, it is enough to prove that the second derivative of $I(P_X\times[\overline{\alpha}\cdot Q_Z +\alpha \cdot V_{Z|X}^* ])$ w.r.t.\ $\alpha\in(0,1)$ is not equal to zero. Now, 
\begin{flalign}
&\frac{d^2}{d\alpha^2}I(P_X\times[\overline{\alpha}\cdot Q_Z +\alpha \cdot V_{Z|X}^* ])\nonumber\\
&=\frac{d^2}{d\alpha^2}\sum_{x,z}P(x) \left([\overline{\alpha} Q_Z(z)+\alpha \cdot V^*(z|x)]
\log\frac{[\overline{\alpha}Q_Z(z) +\alpha \cdot V^*(z|x)]}{[\overline{\alpha}Q_Z(z) +\alpha \cdot V^*(z)]} \right).\label{eq: ;isafudgvi;udgf}
\end{flalign}
Denote $f(\alpha)=\overline{\alpha}\cdot Q_Z(z) +\alpha \cdot V_{Z|X}^*(z|x)$ and $g(\alpha)=\overline{\alpha}Q_Z(z) +\alpha \cdot V^*(z)$. Consider the derivative
\begin{flalign}
&\frac{d}{d\alpha}f(\alpha)\log \frac{f(\alpha)}{g(\alpha)}\nonumber\\
&=f'(\alpha)\log f(\alpha)+f'(\alpha)-f'(\alpha)\log g(\alpha)-\frac{f(\alpha)g'(\alpha)}{g(\alpha)},
\end{flalign}
and note that since both $f(\alpha)$ and $g(\alpha)$ are positive (because $Q_Z(z)=\frac{1}{|\calZ|}$ and $\alpha\in(0,1)$) and affine in $\alpha$, we have $f''(\alpha)=g''(\alpha)=0$, and consequently, the second derivative equals
\begin{flalign}
&\frac{d^2}{d\alpha^2}f(\alpha)\log \frac{f(\alpha)}{g(\alpha)}\nonumber\\
&=\frac{[f'(\alpha)]^2}{f(\alpha)}-\frac{f'(\alpha)g'(\alpha)}{g(\alpha)}-
\frac{f'(\alpha)g'(\alpha)g(\alpha)- f(\alpha)[g'(\alpha)]^2}{g^2(\alpha)}\\
&= \frac{[f'(\alpha)g(\alpha)-f(\alpha)g'(\alpha)]^2}{g^2(\alpha)f(\alpha)}.
\end{flalign}

This yields that
\begin{flalign}
&\frac{d^2}{d\alpha^2}I(P_X\times[\overline{\alpha}\cdot Q_Z +\alpha \cdot V_{Z|X}^* ])\nonumber\\
&= \sum_{x,z:\;P(x)[\overline{\alpha} Q_Z(z)+\alpha \cdot V^*(z|x)]>0 }P(x) \frac{[Q(z)(V_{Z|X}^*(z|x)-V_Z^*(z))]^2}{\left(\overline{\alpha}Q(z)+\alpha V_{Z}^*(z)\right)^2\left(\overline{\alpha}Q(z)+\alpha V_{Z|X}^*(z|x)\right)}.
\end{flalign}
Since $Q_Z(z)=\frac{1}{|\calZ|}$, we obtain zero second derivative 
 iff for all $(x,z)$ it holds that $P(x)(V^*(z)-V^*(z|x))=0$; i.e., iff $I(P_X\times V_{Z|X}^*)=0$.

Thus, if $I(P_X\times V_{Z|X}^* )= R>0$ then from (\ref{eq: ads;oh;odivi}) it follows that $I(P_X\times[\overline{\alpha}\cdot Q_Z +\alpha \cdot V_{Z|X}^* ])<\alpha R$. 
Now,
\begin{flalign}
&\min_{V_{Z|X}:\; I(P\times V_{Z|X} )\leq R}
\alpha \cdot D(V_{Z|X} \|W_{Y|X}|P)\nonumber\\
&\triangleq \alpha \cdot D(V^*_{Z|X} \|W_{Y|X}|P)\label{eq: afhviodufhiovhdofiv}\\
&\geq D(\overline{\alpha}\cdot Q_Z +\alpha \cdot V^*_{Z|X} \|W_{Z|X}^{(\alpha)}|P)\label{eq: a;oifhv;oidhfv}\\
&\geq \min_{P_{Z|X}:\; I(P\times P_{Z|X})\leq \alpha R} D(P_{Z|X}\|W_{Z|X}^{(\alpha)}|P),\label{eq: aiufgviudfvfa}
\end{flalign}
where (\ref{eq: a;oifhv;oidhfv}) follows by the convexity of the divergence.
So, by a sandwich argument, if (\ref{eq: afhviodufhiovhdofiv})-(\ref{eq: aiufgviudfvfa}) are in fact equalities, we have that $\overline{\alpha}\cdot Q_Z +\alpha \cdot V^*_{Z|X} $ is a minimizer of $\min_{P_{Z|X}:\; I(P\times P_{Z|X})\leq \alpha R} D(P_{Z|X}\|W_{Z|X}^{(\alpha)}|P)$ which satisfies the constraint $I(P\times P_{Z|X})\leq \alpha R$ with loose inequality, in contradiction to (\ref{eq: aiufhv;iufhd;ivhdfiohv;d}).

To conclude, for all $R>0$ and $\alpha\in(0,1)$, we have from (\ref{eq: aoufhv;ufdhv}) and (\ref{eq: a;iufgviufdgvavh;fd}) that
\begin{flalign}
\overline{E}^q(\alpha R,P,W_{YZ|X}^{(\alpha)})&=\min_{\substack{P_{XZ}:\; P_X=P\\
I(P_{XZ})\leq \alpha R}}\overline{E}^{q}(P_{XZ},W_{YZ|X}^{(\alpha)}) \nonumber\\
&<\alpha \min_{V_{Z|X}:\; I(P_{XZ})\leq R} D(V_{Z|X}\|W_{Y|X}|P)+\overline{\alpha} \cdot \overline{E}_0^q(P,W)\\
&=\alpha E_{sp}(R,P,W)+\overline{\alpha} \cdot \overline{E}_0^q(P,W).
\end{flalign}

\subsection{Proof of Proposition \ref{pr: comarison to previous sp}}\label{sc: comarison to previous sp}
For any given $W_{Z|XY}$, it holds that 
\begin{flalign}
&\overline{E}^{q}(R,P,W_{YZ|X})\triangleq \min_{\substack{P_{XZ}:\\
P_X=P,\\  I(X;Z)\leq R }}
\max_{\substack{P_{U\widetilde{X}|XZ}:\\ \widetilde{X}-(U,Z)-X\\
P_{\widetilde{X}ZU}=P_{XZU}
}}
\min_{ \substack{ P_{Y|UX\widetilde{X}Z}:\\ U-(\widetilde{X},X,Z)-Y,\\
\EE q(\widetilde{X},Y)\geq \EE q(X,Y)
} } 
D(P_{YZ|X}\|W_{YZ|X}|P)+I(\widetilde{X};Y|X,Z)
\nonumber\\
&\leq 
\min_{\substack{P_{XZ}:\\
P_X=P,\\  I(X;Z)\leq R }}
\max_{\substack{P_{U\widetilde{X}|XZ}:\\ \widetilde{X}-(U,Z)-X\\
P_{\widetilde{X}ZU}=P_{XZU}
}}
\min_{ \substack{ P_{Y|UX\widetilde{X}Z}:\\ U-(\widetilde{X},X,Z)-Y,\\
\EE q(\widetilde{X},Y)\geq \EE q(X,Y)\\\widetilde{X}-(X,Z)-Y
} } 
D(P_{YZ|X}\|W_{YZ|X}|P)\label{eq: a;foihv;oidfhv}\\
&=
\min_{\substack{P_{XZ}:\\
P_X=P,\\  I(X;Z)\leq R }}
\min_{  P_{Y|XZ}: P_{Z|X}\times P_{Y|XZ}\in \calW_q(P_X)
} 
D(P_{YZ|X}\|W_{YZ|X}|P)\label{eq: ytytfytfytfy}\\
&=
\min_{\substack{P_{XYZ}:  
P_X=P,\\  I(X;Z)\leq R \\P_{YZ|X}\in \calW_q(P_X)}}
D(P_{YZ|X}\|W_{YZ|X}|P),
\end{flalign}
where (\ref{eq: a;foihv;oidfhv}) holds due to the additional constraint $\widetilde{X}-(X,Z)-Y$, and (\ref{eq: ytytfytfytfy}) follows by definition of $\calW_q(P_X)$ in (\ref{eq: calW q dfnasdfionidaf;i;}). Finally, taking the infimum over $W_{Z|XY}$ on both sides of the inequality, we obtain the desired results since in the r.h.s., $W_{Z|XY}$ can be chosen to equal $P_{Z|XY}$ yielding 

\noindent $ \inf_{W_{Z|XY}} \overline{E}^{q}(R,P,W_{YZ|X})\leq E_{sp}^q(R,P,W)$.

\subsection{Proof of (\ref{eq: idfuhvilufg})}\label{ap: uniformity}
Note that by applying Bayes' law twice we have 
\begin{flalign}
 &\Pr(\bX=\bx_i(\bz,\widehat{P}_{XZ}),\bZ=\bz|\widehat{P}_{\bX\bZ}=\widehat{P}_{XZ})\nonumber\\
&= \Pr(\bZ=\bz|\widehat{P}_{\bX\bZ}=\widehat{P}_{XZ})\cdot  \Pr(\bX=\bx_i(\bz,\widehat{P}_{XZ})|\bZ=\bz,\widehat{P}_{\bX\bZ}=\widehat{P}_{XZ})\\
&= \Pr(\bX=\bx_i(\bz,\widehat{P}_{XZ})|\widehat{P}_{\bX\bZ}=\widehat{P}_{XZ})\cdot \Pr(\bZ=\bz|\bX=\bx_i(\bz,\widehat{P}_{XZ}),\widehat{P}_{\bX\bZ}=\widehat{P}_{XZ}).
\end{flalign}
Now, since the code is constant composition, the actual joint type-class $\widehat{P}_{\bX\bZ}$ does not depend on the codeword $\bX$, and hence $\Pr(\bX=\bx_i(\bz,\widehat{P}_{XZ})|\widehat{P}_{\bX\bZ}=\widehat{P}_{XZ})=\frac{1}{\mathbb{M}_n}$, and further we have $\Pr(\bZ=\bz|\bX=\bx_i(\bz,\widehat{P}_{XZ}),\widehat{P}_{\bX\bZ}=\widehat{P}_{XZ})=\frac{1}{|\calT_n(\widehat{P}_{Z|X})|}$, this yields
\begin{flalign}
  \Pr(\bX=\bx_i(\bz,\widehat{P}_{XZ})|\bZ=\bz,\widehat{P}_{\bX\bZ}=\widehat{P}_{XZ})&=\frac{1}{|\calT_n(\widehat{P}_{Z|X})|\cdot \mathbb{M}_n\cdot \Pr(\bZ=\bz|\widehat{P}_{\bX\bZ}=\widehat{P}_{XZ})}.
\end{flalign}
Since the r.h.s.\ does not depend on $i$ we obtain the desired result (\ref{eq: idfuhvilufg}). 

\subsection{Proof of Lemma \ref{lm: audvgliudgvliu}}\label{sc: fa;ihvoidfhiovhdfv}

Consider the degenerate case where $\bz$ is a null (a constant sequence, $\forall i$, $\bz(i)=c$), and define
for an empirical distribution $P_{\widetilde{X}X}\in\calP_n(\calX^2)$
\begin{flalign}
&\overline{\Omega}^q_n(P_{\widetilde{X}X}, W_{Y|X})\triangleq 
\min_{\substack{V_{\widetilde{X}XY}\in\calP_n(\calX^2\times\calY):\;
V_{\widetilde{X}X}=P_{\widetilde{X}X},\; 
q(V_{\widetilde{X}Y})\geq q(V_{XY})
}} D(V_{Y|X\widetilde{X}}\|W_{Y|X}|P_{X\widetilde{X}})\label{eq: Ialpha dfndfuiugliuggilugi}.
\end{flalign}

Consider the following definition which is in fact equivalent to Definition \ref{df: aidjosj} (see (\ref{eq: idivugfiv})) for the case in which the sequence $\bz$ is null. 
\begin{definition}
Let $\widehat{P}_{X}$ be given. A codebook $\{\bx_i\}\subseteq\calT_n(\widehat{P}_{X})$ of size $M'$ which satisfies \begin{flalign}\label{eq: idivugfivBOB}
|\hat{P}_{\bx_m\bx_j}(x,x')-\hat{P}_{\bx_\ell\bx_k}(x,x')|\leq 6/\sqrt{M'}+ 2\sqrt{2}/\sqrt{t} +3/t
\end{flalign}\noindent for all $i\neq j$ and
$\ell\neq k$ (not necessarily different from $m$ and $j$) and any $(x, x')\in\calX^2$, is said to be \underline{$(t,\widehat{P}_{X})$-quasi-symmetric}.
\end{definition}

In \cite[Eqs.\ (192)-(197)]{BondaschiGuilleniFabregasDalai-IT2021}, it was shown that if a channel-metric pair $(W_{Y|X},q)$ is balanced, then for any code $\calC_n'$ which is $(t,\widehat{P}_{X})$-quasi-symmetric it holds that there exists a constant $\overline{s}\geq 0$ such that
\begin{flalign}
&D_{min}(\calC_n')\nonumber\\
&\triangleq \frac{1}{|\calC_n'|(|\calC_n'|-1)}\sum_{i,j\in\calC_n',\; j\neq i} \min\left\{\overline{\Omega}^q_n(\widehat{P}_{\bx_i\bx_j}, W_{Y|X}),\overline{\Omega}^q_n(\widehat{P}_{\bx_j\bx_i}, W_{Y|X})\right\}\nonumber\\
&\leq
-\frac{1}{n}\frac{|\calC_n'|}{|\calC_n'|-1}
\sum_{t=1}^n
\sum_{\widetilde{x},x}\overline{P}^{\calC_n'}(x|t)\cdot \overline{P}^{\calC_n'}(\widetilde{x}|t)\log \left(\sum_yW_{Y|X}(y|x)e^{\overline{s}[q(\widetilde{x},y)-q(x,y)]}\right)+\zeta_{t,|\calC_n'|}(W),\label{eq: fdihjfpjb}
\end{flalign}
where $\overline{P}^{\calC_n'}(x|t)\triangleq \overline{P}^{\calC_n'}_{X|T}(x|t)$ is the marginal distribution defined in (\ref{eq: iusdgicugd}), and where
\begin{flalign}
\zeta_{t,|\calC_n'|}(W)&= 5K\left(
6/\sqrt{|\calC_n'|}+2\sqrt{2/t}+3/t\right)\\
K&=\max_{s\in[0,\hat{s}]} \sum_{(x,\widetilde{x})\in\calX^2}
|\mu_{x,\widetilde{x}}(s)|\\
\mu_{x,\widetilde{x}}(s)&= -\log \sum_{y:\; q(\widetilde{x},y)>-\infty,q(x,y)>-\infty } W(y|x)e^{s[q(\widetilde{x},y)-q(x,y)]}\\
\hat{s}&= \max_{(x,\widetilde{x})\in\calX^2}\left\{
\argsup_{s\geq 0}\left(\mu_{x,\widetilde{x}}(s)+ 
\mu_{\widetilde{x},x}(s)\right)\right\}<\infty.\label{eq: hat s leq infty}
\end{flalign}

Next, 
denote
\begin{flalign}
\overline{\zeta}_{t,{\delta},|\calB|}(W_{Z|XY}) &\triangleq 5\overline{K}(W_{Z|XY}^{\delta})\cdot \left(
6/\sqrt{|\calB|}+2\sqrt{2/t}+3/t\right),\\
\overline{K}(W_{Z|XY}^{\delta})&=\max_{s\in[0,s^*(W_{Z|XY}^{\delta})]} \sum_{(x,\widetilde{x},z)\in\calX^2\times\calZ}
\left|\overline{\mu}_{x,\widetilde{x},z}(s,W_{Z|XY}^{\delta})\right|\\
\overline{\mu}_{x,\widetilde{x},z}(s,W_{Z|XY}^{\delta})&= -\log \sum_{y:\; q(\widetilde{x},y)>-\infty,q(x,y)>-\infty } {W}^{{\delta}}(y|x,z)e^{s[q(\widetilde{x},y)-q(x,y)]}\label{eq: over-line mu dfn}\\
s^*(W_{Z|XY}^{\delta})&= \max_{(x,\widetilde{x},z)\in\calX^2\times\calZ}
\argsup_{s\geq 0}\left(\overline{\mu}_{x,\widetilde{x},z}(s,W_{Z|XY}^{\delta})+
\overline{\mu}_{\widetilde{x},x,z}(s,W_{Z|XY}^{\delta})\right).\label{eq: aiufgiufghviudf}
\end{flalign}

Using Lemma \ref{lm: Komlos Lemma} above which generalizes \cite[Theorem 3]{BondaschiGuilleniFabregasDalai-IT2021}, and the fact that $({W}^{{\delta}}_{Y|XZ},q)$ is balanced due to (\ref{eq: ahfuv W star}), 
we can apply (\ref{eq: fdihjfpjb}) to $\calB$ in the role of $\calC_n'$, and ${W}^{{\delta}}_{Y|XZ}$ in the role of $W_{Y|X}$, by treating $(x,z)$ and $(\widetilde{x},z)$ as the input symbols, and $(\bx_i,\bz)$ in the role of $\bx_i$, and then taking the supremum over $s\geq 0$, and obtain that
\begin{flalign}
&\frac{1}{|\calB|(|\calB|-1)}\sum_{i,j\in\calB,\; j\neq i}\min\{\Omega^q_n(\widehat{P}_{\bx_i\bz\bx_j}, {W}^{{\delta}}_{Y|XZ}),
\Omega^q_n(\widehat{P}_{\bx_j\bz\bx_i}, {W}^{{\delta}}_{Y|XZ})\}
\\
&\leq \sup_{s\geq 0}
-\frac{1}{n}\frac{|\calB|}{|\calB|-1}\sum_{t=1}^n \sum_{\widetilde{x},x,z}
\overline{P}^{\calB}_{XZ}(x,z|t)\overline{P}^{\calB}_{XZ}(\widetilde{x},z|t)
\log \left(\sum_y {W}^{{\delta}}_{Y|XZ}(y|x,z)e^{s[q(\widetilde{x},y)-q(x,y)]}\right)\nonumber\\
&+ \zeta_{t,|\calB|}({W}^{{\delta}}_{Y|XZ})\\
&= \sup_{s\geq 0}
-\frac{|\calB|}{|\calB|-1}\sum_{\widetilde{x},x,z}
\overline{P}^{\calB}_{XZ\widetilde{X}}(x,z,\widetilde{x})
\log \left(\sum_y {W}^{{\delta}}_{Y|XZ}(y|x,z)e^{s[q(\widetilde{x},y)-q(x,y)]}\right)+\zeta_{t,|\calB|}({W}^{{\delta}}_{Y|XZ})\label{eq: fiuhviozudfhsovz;ihdfo;ivho;idfv}\\
&= \frac{|\calB|}{|\calB|-1}\cdot \Omega^q\left(\overline{P}^{\calB}_{XZ\widetilde{X}},  {W}^{{\delta}}_{Y|XZ}\right)+\zeta_{t,|\calB|}({W}^{{\delta}}_{Y|XZ})
.
\end{flalign}
Note that $ \zeta_{t,|\calB|}({W}^{{\delta}}_{Y|XZ})$ is in fact a function of $(t,|\calB|,W_{Z|XY},{\delta})$.  Hence, denote $\overline{\zeta}_{t,{\delta},|\calB|}(W_{Z|XY})=\zeta_{t,|\calB|}({W}^{{\delta}}_{Y|XZ})$.

Next, 
we prove that $\overline{\zeta}_{t,{\delta},|\calB|}(W_{Z|XY})$ can be uniformly bounded for every $W_{Z|XY}$. 
To this end, note that (\ref{eq: ahfuv W starert7eft}) and (\ref{eq: ahfuadvdvdsv W starert7eft}) imply that for any $W_{Z|XY}$ it holds that for all $(\widetilde{x},x,s)$
\begin{flalign}
&\left|\mu_{x,\widetilde{x}}(s)- \overline{\mu}_{x,\widetilde{x},z}(s,W_{Z|XY}^{\delta})\right|\leq \log \frac{|\calZ|}{{\delta}}\\
&\left|\left(\mu_{x,\widetilde{x}}(s)+ 
\mu_{\widetilde{x},x}(s)\right)-\left(\overline{\mu}_{x,\widetilde{x},z}(s,W_{Z|XY}^{\delta})+ 
\overline{\mu}_{\widetilde{x},x,z}(s,W_{Z|XY}^{\delta})\right)\right|\leq 2\log \frac{|\calZ|}{{\delta}}.
\end{flalign}
From (\ref{eq: hat s leq infty}) and the concavity of $\mu_{\widetilde{x},x}(s)$ (see \cite[Lemma 1]{BondaschiGuilleniFabregasDalai-IT2021}), this implies that there exists a constant $\theta({\delta})<\infty$, independent of $W_{Z|XY}$ such that $s^*(W_{Z|XY}^{\delta})$ defined in (\ref{eq: aiufgiufghviudf}) satisfies 
\begin{flalign}
s^*(W_{Z|XY}^{\delta})\leq \hat{s}+\theta({\delta})<\infty.
\end{flalign}
Therefore, 
\begin{flalign}
\overline{K}(W_{Z|XY}^{\delta})&=\max_{s\in[0,s^*(W_{Z|XY}^{\delta})]} \sum_{(x,\widetilde{x},z)\in\calX^2\times\calZ}
\left|\overline{\mu}_{x,\widetilde{x},z}(s,W_{Z|XY}^{\delta})\right|\nonumber\\
&\leq \max_{s\in[0,\hat{s}+\theta({\delta})]} \sum_{(x,\widetilde{x},z)\in\calX^2\times\calZ}
\left(\left|\mu_{x,\widetilde{x}}(s)\right|+\left|\log \frac{|\calZ|}{{\delta}}
\right|\right)\nonumber\\
&\triangleq \widetilde{K}({\delta}), 
\end{flalign}
and consequently
\begin{flalign}
\overline{\zeta}_{t,{\delta},|\calB|}(W_{Z|XY}) &\leq 5\widetilde{K}({\delta})\cdot \left(
6/\sqrt{|\calB|}+2\sqrt{2/t}+3/t\right).
\end{flalign}

\subsection{Proof of Lemma \ref{lm: Caratheodory}}\label{ap: Caratheodory}
We prove that while $T$ is a random variable uniformly distributed over $\{1,...,n\}$, indeed the cardinality of the alphabet of the random variable $U$ can
be limited without loss of generality by $|\calX|^2\cdot |\calZ|$. 
This is done by an application of
the support Lemma (Caratheodory's Theorem).
Note that by Bayes' Law $\overline{P}^{\calG_k}_{X\widetilde{X}Z}(x,\widetilde{x},z)= \sum_t \overline{P}^{\calG_k}_T(t) \frac{\overline{P}^{\calG_k}_{ZX|T}(z,x|t)\overline{P}^{\calG_k}_{ZX|T}(z,\widetilde{x}|t)}{V_{Z|T}(z|t)}$, and thus
there exists a distribution $V_{UXZ}$ such that the expectations (w.r.t.\ $U$) of the following $|\calU|\leq |\calX|^2\cdot |\calZ|$ functionals of $V_{XZ|U=u}$:
\begin{flalign}
&\left\{ \frac{V_{ZX|U}(z,x|u)V_{ZX|U}(z,\widetilde{x}|u)}{V_{Z|U}(z|u)}\right\}_{(z,x,\widetilde{x})\in\calZ  \times\calX^2 }, 
\end{flalign}
preserve those of $\overline{P}^{\calG_k}_{TXZ}$; i.e., 
\begin{flalign}
\forall (z,x,\widetilde{x}),\; \sum_u V_U(u) \frac{V_{ZX|U}(z,x|u)V_{ZX|U}(z,\widetilde{x}|u)}{V_{Z|U}(z|u)}&= 
\sum_t \overline{P}^{\calG_k}_{T} (t)\frac{\overline{P}^{\calG_k}_{ZX|T}(z,x|T=t) \overline{P}^{\calG_k}_{ZX|T}(z,\widetilde{x}|T=t)}{\overline{P}^{\calG_k}_{Z|T}(z|T=t)}
\end{flalign}
because there are in fact $|\calZ | \cdot|\calX|^2-1$ degrees of freedom in $\overline{P}^{\calG_k}_{ZX\widetilde{X}}(z,x,\widetilde{x})$, it suffices to preserve only $|\calX|^2\cdot |\calZ|-1$ of the functionals. 
 
As for the last assertion of the lemma, note that preserving the expectation of one additional functional $H_V(Z|U=u) =-\sum_{x,z}V_{ZX|U}(z,x|u)\log \sum_{x'}V_{ZX|U}(z,x'|u)$ yields 
\begin{flalign}
\sum_u V_U(u) H_{\overline{T}}(Z|U=u) &= 
\sum_t \overline{P}^{\calG_k}_{T} (t)
H_{\overline{P}^{\calG_k}_{Z|T}}(Z|T=t),
\end{flalign}
since $H_{\overline{P}^{\calG_k}}(Z|T) =0$, $Z$ is also  deterministic function of $U$, where the alphabet cardinality increase is $1$.

\subsection{Proof of Lemma \ref{eq: List size lemma}}\label{sc: LIST  size lemma Proof}
From the law of total probability
\begin{flalign}
&\Pr\left(
 |\calL(\bZ,\widehat{P}_{XZ})|<e^{n\tau}\big|\widehat{P}_{\bX\bZ}=\widehat{P}_{XZ}\right)\\
&=\frac{1}{\mathbb{M}_n}\sum_{i=1}^{\mathbb{M}_n} \Pr\left(
 |\calL(\bZ,\widehat{P}_{XZ})|<e^{n\tau}\big|\bX=\overline{\bx}_i,\widehat{P}_{\bX\bZ}=\widehat{P}_{XZ}\right)\label{eq: X1}\\
&=\frac{1}{\mathbb{M}_n}\sum_{i=1}^{\mathbb{M}_n} \left|\{\bz\in \calT_n(\widehat{P}_{Z|X}|\overline{\bx}_i):\;|\calL| <e^{n\tau}\}\right|\cdot \frac{1}{|\calT_n(\widehat{P}_{Z|X}|\overline{\bx}_i)|}\label{eq: X2}\\
&=\frac{1}{\mathbb{M}_n}\sum_{\bz\in\calT_n(\widehat{P}_Z):\; |\calL| <e^{n\tau}} |\calL|\cdot \frac{1}{|\calT_n(\widehat{P}_{Z|X}|\overline{\bx}_i)|}\label{eq: X4}\\
&\leq\frac{1}{\mathbb{M}_n}\cdot e^{n\tau} \cdot |\calT_n(\widehat{P}_Z)| \cdot \frac{1}{|\calT_n(\widehat{P}_{Z|X}|\overline{\bx}_i)|}\label{eq: X3}\\
&\leq (n+1)^{|\calX||\calZ|-1}\cdot e^{-n[R-I(\widehat{P}_{XZ})-\tau]}\label{eq: X5}\\
&=e^{-n[R-I(\widehat{P}_{XZ})-\tau-\frac{|\calX||\calZ|-1}{n}\log(n+1)]},
\end{flalign}
where (\ref{eq: X2}) follows since $\bZ$ is uniform over $\calT_n(\widehat{P}_{Z|X}|\overline{\bx}_i)$ given $\overline{\bx}_i$, (\ref{eq: X3}) holds by replacing the count over codewords by a count over sequences $\bz$, and (\ref{eq: X5}) follows by a standard bound on the size of a type-class.

\subsection{Proof of Lemma \ref{lm: continuity sym result}}\label{sc: Proof of continuity sym lemma}

Recall the notation (already defined in (\ref{eq: over-line mu dfn}))
\begin{flalign}
\overline{\mu}_{x,\widetilde{x},z}(s,\overline{W}_{Z|XY})&= -\log \sum_{y:\; q(\widetilde{x},y)>-\infty,q(x,y)>-\infty }\overline{W}(y|x,z)e^{s[q(\widetilde{x},y)-q(x,y)]}
\end{flalign}

Note that for $P_{\widetilde{X}XZ}\in \calP_{sym}(\calX^2\times \calZ)$ 
\begin{flalign}
&\Omega^q(P_{XZ\widetilde{X}},\overline{W}_{Y|XZ})\nonumber\\
&= \sup_{s\geq 0}\sum_{\widetilde{x},x,z}P_{XZ\widetilde{X}}(x,z,\widetilde{x})
\overline{\mu}_{x,\widetilde{x},z}(s,\overline{W}_{Z|XY})
\label{eq: aifuihiovdhfi;o}\\
&= \sup_{s\geq 0}\frac{1}{2}\sum_{\widetilde{x},x,z}P_{XZ\widetilde{X}}(x,z,\widetilde{x})
\left(\overline{\mu}_{x,\widetilde{x},z}(s,\overline{W}_{Z|XY})+\overline{\mu}_{\widetilde{x},x,z}(s,\overline{W}_{Z|XY})\right),
\end{flalign}
where (\ref{eq: aifuihiovdhfi;o}) follows from (\ref{eq: afovfihosbjpfogjbposjgfbojgojfpjb}), and the last step follows from symmetry. 

Since $(q,\overline{W}_{Y|XZ})$ is balanced, from \cite[Lemma 6, Eq.\ (144)]{BondaschiGuilleniFabregasDalai-IT2021} it follows that
\begin{flalign}
&\argsup_{s\geq 0}\frac{1}{2}\sum_{\widetilde{x},x,z}P_{XZ\widetilde{X}}(x,z,\widetilde{x})
\left(\overline{\mu}_{x,\widetilde{x},z}(s,\overline{W}_{Z|XY})+\overline{\mu}_{\widetilde{x},x,z}(s,\overline{W}_{Z|XY})\right)<\infty
\end{flalign}
Let $P_{XZ\widetilde{X}}^{(1)},P_{XZ\widetilde{X}}^{(2)}$ be such that $\forall (x,z,\widetilde{x})$, $|P_{XZ\widetilde{X}}^{(1)}(x,z,\widetilde{x})-P_{XZ\widetilde{X}}^{(2)}(x,z,\widetilde{x})|\leq \epsilon $, 
this implies that there exist finite $s_1,s_2$ such that 
\begin{flalign}
&\Omega^q(P_{XZ\widetilde{X}}^{(1)},\overline{W}_{Y|XZ})-\Omega^q(P_{XZ\widetilde{X}}^{(2)},\overline{W}_{Y|XZ})\nonumber\\
&=
\sum_{\widetilde{x},x,z}P_{XZ\widetilde{X}}^{(1)}(x,z,\widetilde{x})\left(\overline{\mu}_{x,\widetilde{x},z}(s_1,\overline{W}_{Z|XY})+\overline{\mu}_{\widetilde{x},x,z}(s_1,\overline{W}_{Z|XY})\right)\\
&-
\sum_{\widetilde{x},x,z}P_{XZ\widetilde{X}}^{(2)}(x,z,\widetilde{x})\left(\overline{\mu}_{x,\widetilde{x},z}(s_2,\overline{W}_{Z|XY})+\overline{\mu}_{\widetilde{x},x,z}(s_2,\overline{W}_{Z|XY})\right)\\
&\leq 
\sum_{\widetilde{x},x,z}P_{XZ\widetilde{X}}^{(1)}(x,z,\widetilde{x})\left(\overline{\mu}_{x,\widetilde{x},z}(s_1,\overline{W}_{Z|XY})+\overline{\mu}_{\widetilde{x},x,z}(s_1,\overline{W}_{Z|XY})\right)\\
&-
\sum_{\widetilde{x},x,z}P_{XZ\widetilde{X}}^{(2)}(x,z,\widetilde{x})\left(\overline{\mu}_{x,\widetilde{x},z}(s_1,\overline{W}_{Z|XY})+\overline{\mu}_{\widetilde{x},x,z}(s_1,\overline{W}_{Z|XY})\right)\label{eq: sabra}\\
&\leq \epsilon \cdot 
\sum_{\widetilde{x},x,z}\left(\overline{\mu}_{x,\widetilde{x},z}(s_1,\overline{W}_{Z|XY})+\overline{\mu}_{\widetilde{x},x,z}(s_1,\overline{W}_{Z|XY})\right),
\end{flalign}
where (\ref{eq: sabra}) follows since $s_2$ is the argsup of the subtracted term. 

Switching roles between $P_{XZ\widetilde{X}}^{(1)}$ and $P_{XZ\widetilde{X}}^{(2)}$ we obtain
\begin{flalign}
&\left|\Omega^q(P_{XZ\widetilde{X}}^{(1)},\overline{W}_{Y|XZ})-\Omega^q(P_{XZ\widetilde{X}}^{(2)},\overline{W}_{Y|XZ})\right|\nonumber\\
&\leq  \epsilon \cdot \max_{i\in\{1,2\}}
\sum_{\widetilde{x},x,z}\left(\overline{\mu}_{x,\widetilde{x},z}(s_i,\overline{W}_{Z|XY})+\overline{\mu}_{\widetilde{x},x,z}(s_i,\overline{W}_{Z|XY})\right).
\end{flalign}

\subsection{Proof of Lemma \ref{lm: technical ahuv;oiuhdfiouh}}\label{sc: Proof tec}

Denote $\calK=\{1,...,|\calX|^2|\calZ|\}$. 
Let $P_{XZ}^*\in\calP(\calX\times \calZ):\; P_{X}=P_n,\;
I(P_{XZ})\leq R$ be given\footnote{In the sequel we shall prove that the infimum is attained, so $P_{XZ}^*$ can be thought of as a minimizer rather than approximating the infimum.} that approximates the following infimum by , say $1/n$; i.e., 
\begin{flalign}
&
D(P_{Z|X}^*\| {W}^{{\delta}}_{Z|X}|P_X)+\max_{P_{U|XZ}\in\calP(\calK|\calX\times\calZ)}\widetilde{\Omega^q}(P^*_{XZ},P_{U|XZ},  {W}^{{\delta}}_{Y|XZ})\nonumber\\
&\leq \inf_{\substack{
P_{XZ}\in\calP(\calX\times \calZ):\; P_{X}=P_n,\\
I(P_{XZ})\leq R}}D(P_{Z|X}\| {W}^{{\delta}}_{Z|X}|P_X)+\max_{P_{U|XZ}}\widetilde{\Omega^q}(P_{XZ},P_{U|XZ},  {W}^{{\delta}}_{Y|XZ})+\frac{1}{n}.
\end{flalign}
Let $\calB=\{1,...,|\calX|\cdot 2^{|\calZ|}\}$. 
Note that 
\begin{flalign}
&
D(P_{Z|X}^*\| {W}^{{\delta}}_{Z|X}|P_X)+\max_{P_{U|XZ}\in\calP(\calK|\calX\times\calZ)}\widetilde{\Omega^q}(P^*_{XZ},P_{U|XZ},  {W}^{{\delta}}_{Y|XZ})\nonumber\\
&=
D(P_{Z|X}^*\| {W}^{{\delta}}_{Z|X}|P_X)+\max_{P_{U|XZ}\in\calP(\calK\times \calB|\calX\times\calZ)}\widetilde{\Omega^q}(P^*_{XZ},P_{U|XZ},  {W}^{{\delta}}_{Y|XZ})\label{eq: aliufgvliugviughf}
\end{flalign}
where (\ref{eq: aliufgvliugviughf}) follows from Lemma \ref{lm: Caratheodory}, i.e., it is sufficient to take the alphabet cardinality of $U$ to be $|\calK|=|\calX|^2|\calZ|$, but for technical reasons we allow it be bigger. 

We use \cite[Lemma 10]{SomekhBaruchArxiv_16March2022} to express $P_{XZ}^*=P_n\times P_{Z|X}^*$ as a convex combination of empirical distributions, with $(\calX,\calZ,P_{XZ},n)$ in the roles of $(\calA,\calB,P_{AB},\ell)$, respectively, to obtain 
 \begin{flalign}
P_{XZ}^*&= \sum_i \beta_i \cdot P^{(i)}_{XZ}
\end{flalign}
where 
\begin{flalign}
&\forall i,\; P^{(i)}_{X}=P_X=P_n,\; \forall (x,z),\; |P_{XZ}^*(x,z)-P_{XZ}^{(i)}(x,z)|\leq\frac{1}{n}\label{eq: afihvofidhvafdvf}\\
&\|P_{XZ}^*-P_{XZ}^{(i)}\|\leq\frac{|\calX||\calZ|}{n},\; \mbox{ and }
P^*_{Z|X}(z|x)=0\Rightarrow P_{Z|X}^{(i)}(z|x)=0
.\label{ADDeq: ahv;oifdh;iv}
\end{flalign} 

Now, fix $P_{V|XZ}\in\calP(\calK|\calX\times\calZ)$, and let 
\begin{flalign}
P_{XZV}^{(i)}&=P_n\times P_{Z|X}^{(i)}P_{V|XZ}\\
P_{\widetilde{X}XZV}^{(i)}(\widetilde{x},x,z,v)&= P_{XZV}^{(i)}(x,z,v)P_{X|ZV}^{(i)}(\widetilde{x}|z,v).
\end{flalign}

From (\ref{eq: afihvofidhvafdvf}) we have that for all $(x,z)\in\calX\times\calZ$, it holds that for all $i,j$, $|P_{XZ}^{(i)}(x,z)-P_{XZ}^{(j)}(x,z)|\leq \frac{1}{n}\triangleq \epsilon_{1,n}$.
\begin{lemma}\label{lm: adh;idfhiovhf}
For any $x,\widetilde{x},z$ it holds that 
\begin{flalign}\label{eq: a;iufghviufgvuifgv}
&\left|P_{XZ\widetilde{X}}^{(j)}(x,z,\widetilde{x})-P_{XZ\widetilde{X}}^{(i)}(x,z,\widetilde{x})\right|\leq c(1/\sqrt{n}).
\end{flalign}
where $c(1/\sqrt{n})\triangleq 
\frac{1+|\calX|+|\calV|}{\sqrt{n}}+\frac{2|\calX||\calV|}{n}$.
\end{lemma}
\begin{proof}

Fix $\tilde{\delta}_n>0$. 
Note that if $P_{VZ}^{(i)} (v,z) >\tilde{\delta}_n $, then
\begin{flalign}
&\frac{P_{VZ}^{(j)} (v,z)}{P_{VZ}^{(i)} (v,z)}\nonumber\\
&=\frac{\sum_{x'}P_{XZ}^{(j)} (x',z)P_{V|XZ}(v|x',z)}{P_{VZ}^{(i)} (v,z)
}\\
&\leq \frac{\sum_{x'}[P_{XZ}^{(i)} (x',z)+\epsilon_{1,n}]P_{V|XZ}(v|x',z)}{P_{VZ}^{(i)} (v,z)
}\\
&\leq 1+\frac{\epsilon_{1,n}|\calX|}{\tilde{\delta}_n}.
\end{flalign}
Consequently,
\begin{flalign}
&P_{XZ\widetilde{X}}^{(i)}(x,z,\widetilde{x})\nonumber\\
&= \sum_v P_{VZ}^{(i)} (v,z) P_{\widetilde{X}X|VZ}^{(i)}(\widetilde{x},x|v,z)\\
&\leq \tilde{\delta}_n+ 
 \sum_{v:\;P_{VZ}^{(i)} (v,z) >\tilde{\delta}_n } P_{VZ}^{(i)} (v,z)P_{\widetilde{X}X|VZ}^{(i)}(\widetilde{x},x|v,z)\\
&= \tilde{\delta}_n+ 
\sum_{v:\;P_{VZ}^{(i)} (v,z) >\tilde{\delta}_n } P_{XZ}^{(i)}(x) P_{XZ}^{(i)}(\widetilde{x})\frac{ P_{V|XZ}(x|v,z) P_{V|XZ}(\widetilde{x}|v,z)}{P_{VZ}^{(i)} (v,z)}\\
&\leq  \tilde{\delta}_n+ 
\sum_{\substack{v:\;P_{VZ}^{(i)} (v,z) >\tilde{\delta}_n \\P_{VZ}^{(j)} (v,z)\neq 0}} [P_{XZ}^{(j)}(x) +\epsilon_{1,n}][P_{XZ}^{(j)}(\widetilde{x})+\epsilon_{1,n}]\frac{ P_{V|XZ}(x|v,z) P_{V|XZ}(\widetilde{x}|v,z)}{P_{VZ}^{(j)} (v,z)}\cdot\frac{P_{VZ}^{(j)} (v,z)}{P_{VZ}^{(i)} (v,z)}
\nonumber\\
&+ \sum_{v':\; P_{VZ}^{(j)} (v,z)=0}P_{VZ}^{(i)} (v,z)
\\
&\leq  \tilde{\delta}_n+\left(1+ \frac{\epsilon_{1,n}|\calX|}{\tilde{\delta}_n}\right)
\sum_{v} \left[P_{XZ}^{(j)}(x) P_{XZ}^{(j)}(\widetilde{x})
+\epsilon_{1,n}(P_{XZ}^{(j)}(x)+ P_{XZ}^{(j)}(\widetilde{x}))
\right]\frac{ P_{V|XZ}(x|v,z) P_{V|XZ}(\widetilde{x}|v,z)}{P_{VZ}^{(j)} (v,z)}\nonumber\\
&+ |\calX||\calV|\epsilon_{1,n}
\\
&=  \tilde{\delta}_n+\left(1+ \frac{\epsilon_{1,n}|\calX|}{\tilde{\delta}_n}\right)\left(1+ \frac{\epsilon_{1,n}|\calV|}{\tilde{\delta}_n}\right)
P_{XZ\widetilde{X}}^{(j)}(x,z,\widetilde{x})+ |\calX||\calV|\epsilon_{1,n}
.
\end{flalign}
Switching the roles between $P_{XZ\widetilde{X}}^{(j)}$ and $P_{XZ\widetilde{X}}^{(i)}$ in the above derivation, we obtain also $P_{XZ\widetilde{X}}^{(j)}(x,z,\widetilde{x})\leq \tilde{\delta}_n+\left(1+ \frac{\epsilon_{1,n}|\calX|}{\tilde{\delta}_n}\right)\left(1+ \frac{\epsilon_{1,n}|\calV|}{\tilde{\delta}_n}\right)
P_{XZ\widetilde{X}}^{(i)}(x,z,\widetilde{x})+ |\calX||\calV|\epsilon_{1,n}$, and choosing $\tilde{\delta}_n= \frac{1}{\sqrt{n}}$, we obtain 
(\ref{eq: a;iufghviufgvuifgv}). 
\end{proof}

Next, Let $B$ denote the RV that stands for the index $i$ and satisfies $\Pr(B=i)=\beta_i$. Denoting 
$P_{VBXZ}(v,i,x,z)=\beta_i P_{XZ}^{(i)}(x,z)P_{V|XZ}(v|x,z)$, and further 

\noindent $P^{(i)}_{VXZ\widetilde{X}}(v,x,z,\widetilde{x})=
P_{VXZ|B}(v,x,z|i)P_{X|VBZ}(\widetilde{x}|v,i,z)$, we get that $P_{XZ}=\sum_i\beta_iP_{XZ}^{(i)}=P_{XZ}^*$. Further, let $P_{Y|XZ\widetilde{X}}^*$ denote the minimizer of $\widetilde{\Omega^q}(P_{XZ}^*,P_{V,B|XZ},  {W}^{{\delta}}_{Y|XZ})$. 

Denote by $i_0$ an index which satisfies $\EE_{P_{XZV\widetilde{X}}^{(i_0)}\times P_{Y|XZ\widetilde{X}}^* }[q(\widetilde{x},Y)-q(X,Y)]\geq \EE_{\sum_{i}\beta_i P_{XZV\widetilde{X}}^{(i)}\times P_{Y|XZ\widetilde{X}}^* }[q(\widetilde{x},Y)-q(X,Y)]$, and let $b_\delta(1/\sqrt{n})\triangleq c(1/\sqrt{n})
\log\left(\frac{|\calZ|}{{\delta}\cdot w_{min}}\right)$, where 
\begin{flalign}
w_{min}&\triangleq \min_{(x,y):\; W(y|x)>0} W_{Y|X}(y|x).\label{eq: W min dfn 1} 
\end{flalign}
We get
\begin{flalign}
&\max_{P_{U|XZ}\in\calP(\calK|\calX\times\calZ)} \widetilde{\Omega^q}(P_{XZ}^*,P_{U|XZ},  {W}^{{\delta}}_{Y|XZ})\nonumber\\
&\geq  \widetilde{\Omega^q}(P_{XZ}^*,P_{V,B|XZ},  {W}^{{\delta}}_{Y|XZ})\\
&= \sum_{i,v,x,\widetilde{x},z,y}\beta_i P_{XZV\widetilde{X}}^{(i)}(x,z,v,\widetilde{x})P_{Y|XZ\widetilde{X}}^*(y|x,z,\widetilde{x})
\log\frac{P_{Y|XZ\widetilde{X}}^*(y|x,z,\widetilde{x})}{ {W}^{{\delta}}_{Y|XZ}(y|x,z)}\\
&\geq  \sum_{v,x,\widetilde{x},z,y} P_{XZV\widetilde{X}}^{(i_0)}(x,z,v,\widetilde{x})P_{Y|XZ\widetilde{X}}^*(y|x,z,\widetilde{x})
\log\frac{P_{Y|XZ\widetilde{X}}^*(y|x,z,\widetilde{x})}{ {W}^{{\delta}}_{Y|XZ}(y|x,z)}-
b_\delta(1/\sqrt{n})\label{eq: aiuhv;iufdgviudfhivhdf}
\\
&\geq \widetilde{\Omega^q}(P_{XZ}^{(i_0)},P_{V|XZ},  {W}^{{\delta}}_{Y|XZ})-b_\delta(1/\sqrt{n}),
\end{flalign}
where (\ref{eq: aiuhv;iufdgviudfhivhdf}) follows by the assumption of the lemma that 
${W}^{{\delta}}_{Z|XY}(z|x,y)\geq \frac{{\delta}}{|\calZ|}$, 
which implies that ${W}^{{\delta}}_{Y|XZ}(y|x,z)\geq \frac{{\delta}}{|\calZ|}W_{Y|X}(y|x)$ (\ref{eq: ahfuv W star}), 
yielding $\bigg|\sum_yP^*(y|x,z,\widetilde{x})
\log\frac{P^*(y|x,z,\widetilde{x})}{ {W}^{{\delta}}_{Y|XZ}(y|x,z)}\bigg|\leq 
\log\left(\frac{|\calZ|}{{\delta}\cdot w_{min}}\right)$, and
from Lemma \ref{lm: adh;idfhiovhf} (see (\ref{eq: a;iufghviufgvuifgv})). The last step follows by definition of $i_0$. 

Since this is true for any $P_{V|XZ}$, we can maximize over $P_{V|XZ}$, to obtain
\begin{flalign}
&\max_{P_{U|XZ}} \widetilde{\Omega^q}(P_{XZ}^*,P_{U|XZ},  {W}^{{\delta}}_{Y|XZ})\nonumber\\
&\geq \max_{P_{V|XZ}}\widetilde{\Omega^q}(P_{XZ}^{(i_0)},P_{V|XZ},  {W}^{{\delta}}_{Y|XZ})-b_\delta(1/\sqrt{n}).
\end{flalign}

Hence, denoting $A=|\calX|^2|\calZ|$, and $c_n=\frac{|\calX|^2|\calZ|}{n}$ from \cite[Lemma 2.7]{CsiszarKorner81}, it follows that 
\begin{flalign}
&|D(P_{Z|X}^*\| {W}^{{\delta}}_{Z|X}|P_X)-D(P_{Z|X}^{(i_0)}\| {W}^{{\delta}}_{Z|X}|P_X)|\nonumber\\
&\leq - 2\cdot c_n\log\frac{c_n}{A}+c_n\log \frac{|\calZ|}{{\delta}}\triangleq e_n,
\end{flalign}
 where 
 $\|I(P_n\times P_{Z|X}^*)-I(P_n\times P_{Z|X}^{(i_0)})\|\leq - 2\cdot c_n\log\frac{c_n}{A}\triangleq e_n'$, this gives
\begin{flalign}
&\inf_{\substack{
P_{XZ}\in\calP(\calX\times \calZ):\; P_{X}=P_n,\\
I(P_{XZ})\leq R}}D(P_{Z|X}\| {W}^{{\delta}}_{Z|X}|P_X)+\max_{P_{U|XZ}}\widetilde{\Omega^q}(P_{XZ},P_{U|XZ},  {W}^{{\delta}}_{Y|XZ})+\frac{1}{n}\nonumber\\
&\geq D(P_{Z|X}^*\| {W}^{{\delta}}_{Z|X}|P_X)+\max_{P_{U|XZ}}\widetilde{\Omega^q}(P^*_{XZ},P_{U|XZ},  {W}^{{\delta}}_{Y|XZ})\label{eq: alvb}\\
&\geq  D(P_{Z|X}^{(i_0)}\| {W}^{{\delta}}_{Z|X}|P_X)+\max_{P_{V|XZ}}\widetilde{\Omega^q}(P_{XZ}^{(i_0)},P_{V|XZ},  {W}^{{\delta}}_{Y|XZ})-b_\delta(1/\sqrt{n})-e_n\\
&\geq \min_{\substack{
P_{XZ}\in\calP_n(\calX\times \calZ):\; P_{X}=P_n,\\
I(P_{XZ})\leq R+e_n'}}D(P_{Z|X}\| {W}^{{\delta}}_{Z|X}|P_X)+\max_{P_{V|XZ}}\widetilde{\Omega^q}(P_{XZ},P_{V|XZ},  {W}^{{\delta}}_{Y|XZ})-b_\delta(1/\sqrt{n})-e_n.
\end{flalign}

It remains to prove that the infimum in the l.h.s.\ of (\ref{eq: alvb}) is attained; i.e., the infimum is in fact a minimum. 
Since the sets $\{P_{XZ}\in\calP(\calX\times \calZ):\; P_{X}=P_n, 
I(P_{XZ})\leq R\}$ and $\calP(\calU|\calX\times\calZ)$ are both closed bounded and convex, this will follow by showing that the function $D(P_{Z|X}\| {W}^{{\delta}}_{Z|X}|P_X)+\widetilde{\Omega^q}(P_{XZ},P_{U|XZ},  {W}^{{\delta}}_{Y|XZ})$ is continuous in $P_{XZ}$. 
Now, $D(P_{Z|X}\| {W}^{{\delta}}_{Z|X}|P_X)$ is clearly continuous in $P_{XZ}$. 
It is also easy to verify that $\widetilde{\Omega^q}(P_{XZ},P_{U|XZ},  {W}^{{\delta}}_{Y|XZ})$ is continuous, because for fixed $P_{U|XZ}$, the distribution $P_{XZ\widetilde{X}}$ is continuous in $P_{XZ}$ (this is proved in Lemma \ref{lm: adh;idfhiovhf}), and further, the function $\widetilde{\Omega^q}(P_{XZ},P_{U|XZ},  {W}^{{\delta}}_{Y|XZ})=\Omega^q(P_{XZ\widetilde{X}}, {W}^{{\delta}}_{Y|XZ})$ is continuous in $P_{XZ\widetilde{X}}$ (see Lemma \ref{lm: continuity sym result}).

\subsection{Proof of Proposition \ref{eq: EB inequality}}\label{sc: EB Inequaliity}

Consider the dual expression 
in (\ref{eq: ofauhvoidfhv}). 
Next, consider a sub-optimal choice for $W_{YZ|X}$ such that $Y-X-Z$ forms a Markov chain; i.e., $W_{YZ|X}=W_{Y|X}\times W_{Z|X}$, and take the infimum over $W_{Z|X}$. This yields
\begin{flalign}
&\inf_{W_{Z|XY}}\overline{E}^q(R,P,W_{YZ|X})\nonumber\\
&\leq \inf_{W_{Z|X}}\overline{E}^q(R,P,W_{Y|X}\times W_{Z|X})\label{eq: afhvidfhivihf}\\
&= \inf_{W_{Z|X}}\min_{\substack{P_{XZ}:P_X=P,\\ I(X;Z)\leq R }}
D(P_{Z|X}\|W_{Z|X}|P)+ 
\max_{P_{U|XZ}}\eta_q (P_{XZU},W_{Y|X})\\
&= E_{B}^q(R,P,W),
\label{eq: ofauhvdavadvsoidfhvadfvdfnkll}
\end{flalign}
where the last step follows by switching the order between the infimum over $W_{Z|X}$ and the minimum over $P_{XZ}$, and then taking $W_{Z|X}=P_{X|Z}$ which gives $D(P_{Z|X}\|W_{Z|X}|P)=0$ and is obviously the minimizer.

In remains to show that $\eta_{q_{\mbox{\tiny{ML}}}} (P_{XZU},W_{Y|X})=\sum_{u,\widetilde{x},x,z}P_{XZU}(x,z,u)P_{X|ZU}(\widetilde{x}|u,z)
\log \frac{1}{\sum_y\sqrt{W(y|x)W(y|\widetilde{x})} }$. To see this, recall that by definition (\ref{eq: aiugvliufdglivugdfuiv}):
\begin{flalign}
& \eta_{q_{\mbox{\tiny{ML}}}}(P_{XZU},W_{Y|X}) \nonumber\\
& =\sup_{s\geq 0}-\sum_{u,\widetilde{x},x,z}P_{XZU}(x,z,u)P_{X|ZU}(\widetilde{x}|u,z)\log \sum_y\left.W(y|x)e^{s[q(\widetilde{x},y)-q(x,y)]}\right|_{q= q_{\mbox{\tiny{ML}}}}\\
& =\sup_{s\geq 0}-\sum_{u,\widetilde{x},x,z}P_{XZU}(x,z,u)P_{X|ZU}(\widetilde{x}|u,z)\log \sum_{\substack{y:\; q(\widetilde{x},y)>-\infty,\\q(x,y)>-\infty\\ W(y|x)>0}}\left.W(y|x)e^{s[q(\widetilde{x},y)-q(x,y)]}\right|_{q= q_{\mbox{\tiny{ML}}}}\label{eq: aiugviudfh}\\
& =\sup_{s\geq 0}-\sum_{u,\widetilde{x},x,z}P_{XZU}(x,z,u)P_{X|ZU}(\widetilde{x}|u,z)\log \sum_{y:\; W(y|x)W(y|\widetilde{x})>0}[W(y|x)]^{1-s}[W(y|\widetilde{x})]^s \label{eq: aiugvligdfuivadfnvdfkv}\\
& =\sup_{s\geq 0}-\sum_{u,\widetilde{x},x,z}P_{XZU}(x,z,u)P_{X|ZU}(\widetilde{x}|u,z)\log \sum_{y:\; W(y|x)W(y|\widetilde{x})>0}\sqrt{W(y|x)W(y|\widetilde{x})} \label{eq: aiugvligdfuiv}\\
& =\sup_{s\geq 0}-\sum_{u,\widetilde{x},x,z}P_{XZU}(x,z,u)P_{X|ZU}(\widetilde{x}|u,z)\log \sum_y\sqrt{W(y|x)W(y|\widetilde{x})} ,\label{eq: a;ifuhv;iufdhv}
\end{flalign}
where 
(\ref{eq: aiugviudfh}) follows since we assume $W(y|x)>0\Rightarrow q(x,y)>-\infty$ (see (\ref{eq: positive capacity})), and since $q_{\mbox{\tiny{ML}}}$ obviously satisfies this condition, and in addition, if $q(\widetilde{x},y)=-\infty$, the contribution of the summand $W(y|x)e^{s[q(\widetilde{x},y)-q(x,y)]}$ equals zero. Further, 
(\ref{eq: aiugvligdfuiv}) follows from the concavity in $s$ (see \cite[Lemma 1]{BondaschiGuilleniFabregasDalai-IT2021}) of
the function $-\log \sum_{y:\; W(y|x)W(y|\widetilde{x})>0}W(y|x)e^{s[q(\widetilde{x},y)-q(x,y)]}$, and since the derivative w.r.t.\ $s$ of the function inside the supremum on the r.h.s.\ of (\ref{eq: aiugvligdfuivadfnvdfkv}) equals
\begin{flalign}
&\frac{d}{ds}\Bigg[ \sum_{u,\widetilde{x},x,z}P_{XZU}(x,z,u)P_{X|ZU}(\widetilde{x}|u,z)\log \sum_{y:\; W(y|x)W(y|\widetilde{x})>0}[W(y|x)]^{1-s}[W(y|\widetilde{x})]^s \Bigg]\nonumber\\
&=\Bigg[ \sum_{u,\widetilde{x},x,z}P_{XZU}(x,z,u)P_{X|ZU}(\widetilde{x}|u,z)
\frac{\sum_{y}W(y|x)\left[\frac{W(y|\widetilde{x})}{W(y|x)}\right]^s\cdot \log\frac{W(y|\widetilde{x})}{W(y|x)}}{\sum_{y'}W(y'|x)\left[\frac{W(y'|\widetilde{x})}{W(y'|x)}\right]^s}
\Bigg]\\
&=\frac{1}{2}\Bigg[ \sum_{u,\widetilde{x},x,z}P_{XZU}(x,z,u)P_{X|ZU}(\widetilde{x}|u,z)
\frac{\sum_{y}W(y|x)\left[\frac{W(y|\widetilde{x})}{W(y|x)}\right]^s\cdot \log\frac{W(y|\widetilde{x})}{W(y|x)}}{\sum_{y'}W(y'|x)\left[\frac{W(y'|\widetilde{x})}{W(y'|x)}\right]^s}
\nonumber\\
&+\sum_{u,x,\widetilde{x},z}P_{XZU}(\widetilde{x},z,u)P_{X|ZU}(x|u,z)
\frac{\sum_{y}W(y|\widetilde{x})\left[\frac{W(y|x)}{W(y|\widetilde{x})}\right]^s\cdot \log\frac{W(y|x)}{W(y|\widetilde{x})}}{\sum_{y'}W(y'|\widetilde{x})\left[\frac{W(y'|x)}{W(y'|\widetilde{x})}\right]^s}\Bigg],
\end{flalign}
where the last step follows by switching the names of the summation variables $x,\widetilde{x}$. Substituting $s=1/2$ we obtain
\begin{flalign}
&\frac{1}{2}\Bigg[ \sum_{u,\widetilde{x},x,z}P_{XZU}(x,z,u)P_{X|ZU}(\widetilde{x}|u,z)
\frac{\sum_{y}\sqrt{W(y|x)W(y|\widetilde{x})}
\cdot \log\frac{W(y|\widetilde{x})}{W(y|x)}}{\sum_{y'}\sqrt{W(y'|x)W(y'|\widetilde{x})}}
\nonumber\\
&+\sum_{u,x,\widetilde{x},z}P_{XZU}(\widetilde{x},z,u)P_{X|ZU}(x|u,z)
\frac{\sum_{y}\sqrt{W(y|\widetilde{x})W(y|x)}\cdot \log\frac{W(y|x)}{W(y|\widetilde{x})}}{\sum_{y'}\sqrt{W(y'|\widetilde{x})W(y'|x)}}\Bigg],
\end{flalign}
which obviously equals zero by symmetry and Bayes law, and since $\log\frac{W(y|x)}{W(y|\widetilde{x})}= -\log\frac{W(y|\widetilde{x})}{W(y|x)}$. Finally, (\ref{eq: a;ifuhv;iufdhv}) follows by our assumption that $C_0(W)=0$ and thus for any $(x,\widetilde{x})$, there exists a common channel output, and thus $\sum_y W(y|x)W(y|\widetilde{x})>0$.

To show that the inequality (\ref{eq: afhvidfhivihf}) can be strict, take $W_{Y|X}=W_{bsc}^p$ which denotes the binary symmetric channel (BSC) with crossover probability $p$, and $q=q_{\mbox{\tiny{ML}}}$, and note that lower bounding the maximization over $P_{U|XZ}$ with, e.g., the choice $U=Z$ 
we get
\begin{flalign}
E_{B}(R,P,W_{bsc(p)})\geq 
\min_{\substack{P_{XZ}\in\calP(\calX\times\calZ):\\
P_X=P,\\  I(X;Z)\leq R }}
 \sum_{\widetilde{x},x,z}P_{XZ}(x,z)P_{X|Z}(\widetilde{x}|z)\log \frac{1}{\sum_y\sqrt{W_{bsc(p)}(y|x)W_{bsc(p)}(y|\widetilde{x})} }.\label{eq: aiduvadfvdfgilfudgvq}
\end{flalign}
The r.h.s.\ of (\ref{eq: aiduvadfvdfgilfudgvq}) is nothing but the above mentioned inaccurate bound of \cite[Theorem 12]{BlahutComposition}. 
It was shown in \cite[Fig.\ 2]{BlahutComposition} (see also the corresponding discussion below Fig.\ 2 therein for the case $p=0.1$) that for a range of low rates, the r.h.s.\ of (\ref{eq: aiduvadfvdfgilfudgvq}) maximized over $P$ exceeds the straight-line bound $E_{sl-sp}(R,W)$ (see (\ref{eq: SLLSP})), and consequently, it holds that for $R>0$, $\max_P E_{B}(R,P,W_{bsc}^{p})> E_{sl-sp}(R,W_{bsc}^{p})$. 
Hence, from the additional strict inequality (\ref{eq: aduvg;diufgv}) we get for $W=W_{bsc}^{p}$: 
$ \overline{E}(R,W_{YZ|X}^{(\alpha_{\mbox{\tiny{R}}} )})<E_{sl-sp}(R,W)<\max_P E_{B}(R,P,W)$.

\subsection{Outline of the derivation of the bound $E_{B}(R,P,W)$ of Eq.\ (\ref{eq: aiduvgilfudgv}) based on \cite{BondaschiDalai2022}}\label{ap: CORRECTING BLAHUT}

Consider the derivation which appears in \cite[Section V]{BondaschiDalai2022} up to Eq.\ (82) therein. Then we make two amendments to conclude the proof that are outlined as follows:
\begin{itemize}
\item 
The expression of Eq.\ (82) is upper bounded using Caratheodory's Theorem similar to Lemma \ref{lm: Caratheodory} and (\ref{eq: aiufgviufdgv})-(\ref{eq: adfhv;iudfhv;iodhf;}) to yield the bound 
$E_{B}^{(n)}(R,P,W)$ 
\begin{flalign}
E_{B}^{(n)}(R,P,W)&\triangleq 
\min_{\substack{P_{XV}\in\calP_n(\calX\times\calV):\\
P_X=P,\\  I(X;V)\leq R }}
 \max_{P_{U|XV}}
 \sum_{u,\widetilde{x},x,v}P_{XVU}(x,v,u)P_{X|UV}(\widetilde{x}|u,v)\log \frac{1}{\sum_y\sqrt{W(y|x)W(y|\widetilde{x})} },\label{eq: aiduvgiiugliugiulfudgv}
 \end{flalign}
 where $U$ is a discrete random variable of a specified bounded alphabet. 
Note that $E_{B}^{(n)}(R,P,W)$ differs from $E_{B}(R,P,W)$ by replacing $\calP(\calX\times\calV)$ by $\calP_n(\calX\times\calV)$.

\item 
Then, one can upper bound the minimization over $\calP_n(\calX\times\calV)$ by one that is performed over the simplex $\calP(\calX\times\calV)$ (with additional $o(1)$ loss and a slightly smaller $R$), similarly to Lemma \ref{lm: technical ahuv;oiuhdfiouh} (which can be simplified in the case). 
\end{itemize}

\subsection{Proof of Corollary \ref{cr: quadratic}}\label{sc: Proof of Corollary Blahut}

Similar to $E_{orth}^{q}(R,P,W_{Y|X})$, define
\begin{flalign}
E_{orth}^{q}(R,P,W_{YZ|X})
&=\min_{\substack{P_{\widetilde{X}XYZ}:
P_{XZ}=P_{\widetilde{X}Z},\\  X-Z-\widetilde{X}, I(X;Z)\leq R \\
\EE q(\widetilde{X},Y)\geq \EE q(X,Y),
} } 
D(P_{YZ|X}\|W_{YZ|X}|P)+I(\widetilde{X};Y|X,Z).
\end{flalign}
The inequality $\overline{E}^{q}(R,P,W_{YZ|X})\geq E_{orth}^{q}(R,P,W_{YZ|X})$ follows by taking either $P_{U|XZ}=P_U$ or $P_{U|XZ}=\indicator_{\{U=Z\}}$ in (\ref{eq: afouhvuidfhv}). Noting that the choice $W_{Z|XY}=P_{Z|XY}$ is a minimizer of $E_{orth}^{q}(R,P,W_{YZ|X})$ yields the left side inequality of (\ref{eq: aiuviufhdiuvhdfiuhviudf}).

To prove the r.h.s.\ inequality of (\ref{eq: aiuviufhdiuvhdfiuhviudf}), 
note that
\begin{flalign}
E_{orth}^{q}(R,P,W_{YZ|X})
&\geq \min_{\substack{P_{\widetilde{X}XYZ}:
P_{XZ}=P_{\widetilde{X}Z},\\  X-Z-\widetilde{X}, I(X;\widetilde{X})\leq R \\
\EE q(\widetilde{X},Y)\geq \EE q(X,Y),
} }
D(P_{YZ|X}\|W_{YZ|X}|P)+I(\widetilde{X};Y,X|Z)+\left|I(X;Z)-R\right|_+\label{eq: a'vofkv}\\
&\geq \min_{\substack{P_{\widetilde{X}XYZ}:
P_{XZ}=P_{\widetilde{X}Z},\\  X-Z-\widetilde{X}, I(X;\widetilde{X})\leq R \\
\EE q(\widetilde{X},Y)\geq \EE q(X,Y),
} }
D(P_{YZ|X}\|W_{YZ|X}|P)+\left|I(\widetilde{X};Y,X,Z)-R\right|_+\label{eq: a'voasdcafkv}\\
&\geq \min_{\substack{P_{\widetilde{X}XY}:
P_{XZ}=P_{\widetilde{X}Z},\\ I(X;\widetilde{X})\leq R \\
\EE q(\widetilde{X},Y)\geq \EE q(X,Y),
} }
D(P_{Y|X}\|W_{Y|X}|P)+\left|I(\widetilde{X};Y,X)-R\right|_+\label{eq: a'vofkvadvdfv}\\
&=E^q_{\mbox{\tiny{CK}}}(R,P,W),
\end{flalign}
where (\ref{eq: a'vofkv}) follows since under $I(X;Z)\leq R$, we have $\left|I(X;Z)-R\right|_+=0$, and further since $ X-Z-\widetilde{X}$ is a Markov chain, and thus $I(X;\widetilde{X}|Z)=0$ and also $ I(X;\widetilde{X})\leq I(X;Z)$. Inequality (\ref{eq: a'voasdcafkv}) follows since if $a\geq 0$, then $a+|b|_+\geq |a+b|_+$, 
 and (\ref{eq: a'vofkvadvdfv}) holds since $D(P_{YZ|X}\|W_{YZ|X}|P)\geq D(P_{Y|X}\|W_{Y|X}|P)$ and $I(\widetilde{X};Y,X,Z)\geq I(\widetilde{X};Y,X)$. 
 
 \subsection{Proof of Corollary \ref{eq: U exepmt corllary}}\label{cs: U exepmt corllary}

Clearly, similar to \cite[Appendix A]{SomekhBaruchArxiv_16March2022}), 
the marginal $P_{\widetilde{X}XZ}$ distribution of $P_{\widetilde{X}XZU}$ where $\widetilde{X}-(U,Z)-X$ and $P_{XZU}=P_{\widetilde{X}ZU}$ is symmetric, and thus satisfies $P_{\widetilde{X}XZ}\in \calP_{sym}(\calX^2\times \calZ)$. 
Secondly, note that any distribution $P_{\widetilde{X}XZU}$ where $\widetilde{X}-(U,Z)-X$ and $P_{XZU}=P_{\widetilde{X}ZU}$ satisfies for any $(x,z)\in \calX\times\calZ$:
\begin{flalign}
P_{\widetilde{X}X|Z}(x,x|z)& = \sum_uP_{U|Z}(u|z)\left[P_{X|UZ}(x|u,z)\right]^2\nonumber\\
&\geq \left[\sum_uP_{U|Z}(u|z)P_{X|UZ}(x|u,z)\right]^2\label{eq: aiugviudgv}\\
&=\left[P_{X|Z}(x|z)\right]^2.
\end{flalign}
where (\ref{eq: aiugviudgv}) follows from Jensen's inequality. The equality (\ref{eq: aifudgviufdgvd}) follows from (\ref{eq: ofauhvoidfhv}).


%
%
\end{document}

%% file: Figure_BC_genie_2rec.tex
	\tikzstyle{block} = [draw, fill=white!20, rectangle, 
	minimum height=3em, minimum width=6em]
	\tikzstyle{genblock} = [draw, fill=white!20, circle]
	
	\tikzstyle{block2} = [draw, fill=white!20, rectangle, 
	minimum height=3em, minimum width=3em]    
	\tikzstyle{sum} = [draw, fill=white!20, circle, node distance=1cm]
	\tikzstyle{input} = [coordinate]
	\tikzstyle{output} = [coordinate]
	\tikzstyle{pinstyle} = [pin edge={to-,thin,black}]
	
	\begin{tikzpicture}[auto, node distance=2cm]
	\node [input, name=input] {};

	\node [block, right of=input] (Encoder) at (-1.5,1) {Encoder} ;
	\node [block, right of=Encoder, node distance=3cm] (channel) at (1.5,1) {Channel} ;
	\draw [thick,->] (Encoder) -- node[name=u] {$\bX$} (channel);
	\node [block, right of=channel, node distance=3cm] (Decoder) at (5.5,1) {$Y$-Decoder};
	\draw [thick,->] (channel) -- node[name=y] {$\bY$}  (Decoder) ;  
	\node [output, right of=Decoder] (output) {};
	\node (M) (input)  at (-2.5,1) {$M$};
	\node (M2) (output)  at (11.5,1) {$\hat{M}$};
	\draw [thick](4.6,-2) --  node[name=z] {$\bZ$} (4.6, 0.4);
	\draw [thick](4.6,-4.5) --   (4.6, -2);
	\draw [thick,->](4.6,-2) --   (7.25, -2);
	\draw [thick,->](4.6,-4.5) --   (7.45, -4.5);
	
	\draw [thick](2.5,-5.5) -- (2.5, 1);
	\draw [thick,->](2.5,-5.5) --  node[name=jdfhv] {} (7.45, -5.5);

	\node [block, right of=channel, node distance=3cm] (Decoder2) at (5.5,-2){$  \mbox{List-Genie}$};
	\node [genblock, below of=Decoder2, node distance=2cm] (TGenie) at (8.5,-3) {Type-Genie};
	\draw [thick,->] (TGenie) -- node[name=TTT] {$\hat{P}_{\bX\bZ}$}  (Decoder2) ; 
	
	\draw [thick,->] (Decoder2) --  node[right of= channel, name=L] {$ \calL=\{\bx_1,...,\bx_{|\calL|}\}$}  (Decoder) ; 

	\draw [draw,thick,->] (input) -- (Encoder);
	
	\draw [thick,->] (Decoder) -- (output);

%
%
%
%

%
	
	\end{tikzpicture}
	
	